\documentclass[useAMS,usenatbib,a4paper]{mn2e}
\usepackage{textcomp}
\usepackage{natbib}
\usepackage{lscape}
\usepackage[section] {placeins}
\usepackage{amsfonts, amssymb, amsmath}
\usepackage{aas_macros}
\usepackage{booktabs, times, varioref ,color, multirow, textcomp, graphicx, gensymb}
\usepackage[
    dvips,
    a4paper=true,
    plainpages=true,
    pdfpagelabels,
    bookmarks=true,
    bookmarksopen=false,
    bookmarksopenlevel=2
    bookmarksnumbered=true,
    bookmarkstype=toc,
    colorlinks=true,
    citecolor=red,
    linkcolor=blue,
    menucolor=green,
    urlcolor=magenta,
]{hyperref}
\usepackage[T1]{fontenc}
\usepackage{aecompl} 

\include{aas_journals}

\newcommand{\vect}[1]{\boldsymbol{#1}}
\newcommand{\dd}{\mathrm{d}}
\newcommand{\e}{\mathrm{e}}

\newcommand{\msun}{h^{-1}{\rm M}_{\odot}}
\newcommand{\lgpc}{\,h^{-1}{\rm Gpc}}
\newcommand{\lmpc}{\,h^{-1}{\rm Mpc}}

\newcommand{\kms}{{\rm km\ s^{-1}}}

\newcommand{\mxxl}{{MXXL}}
\newcommand{\cobe}{{\em COBE}}
\newcommand{\wmap}{{\em WMAP}}
\newcommand{\plk}{{\em Planck}}

\newcommand{\healpix}{{\tt HEALPix}}

\newcommand\plotone[1]
 {\centering \leavevmode \includegraphics[width={0.99\columnwidth}]{#1}}

\newcommand{\plotside}[1]
 {\centering \leavevmode \includegraphics[width={0.95\textwidth}]{#1}}
\newcommand{\plotsiderotate}[1]
 {\centering \leavevmode \includegraphics[angle=90,width={0.95\textwidth}]{#1}}

\begin{document}

\title[Optimization of kSZ measurements]
{
Matched filter optimization of kSZ measurements with a reconstructed cosmological flow field.
}
\author[Li et al]
{Ming Li$^{1,2}$\thanks{E-mail:~mingli@mpa-garching.mpg.de, mingli@pmo.ac.cn}, \,R. E. Angulo$^3$, \,S. D. M. White$^2$, \,J. Jasche$^4$  \\
$^1$Purple Mountain Observatory, West Beijing Rd. 2, 210008 Nanjing, People's Republic of China. \\
$^2$Max-Planck-Institute for Astrophysics, Karl-Schwarzschild-Str. 1, D-85740 Garching, Germany. \\
$^3$Centro de Estudios de F\'isica del Cosmos de Arag\'on,  Plaza San Juan 1, Planta-2, 44001, Teruel, Spain. \\
$^4$Institut d'Astrophysique de Paris (IAP), UMR 7095, CNRS - UPMC Universit\'e Paris 6,\\ 98bis boulevard Arago, F-75014 Paris, France.}


\date{\today}

\pagerange{\pageref{firstpage}--\pageref{lastpage}}
\pubyear{2014}

\maketitle
\label{firstpage}

\begin{abstract}
We develop and test a new statistical method to measure the kinematic
Sunyaev-Zel'dovich (kSZ) effect. A sample of independently detected clusters is
combined with the cosmic flow field predicted from a galaxy redshift survey in
order to derive a matched filter that optimally weights the kSZ signal for the
sample as a whole given the noise involved in the problem. We apply this
formalism to realistic mock microwave skies based on cosmological $N$-body
simulations, and demonstrate its robustness and performance. In particular, we
carefully assess the various sources of uncertainty, cosmic microwave background primary fluctuations,
instrumental noise, uncertainties in the determination of the velocity field,
and effects introduced by miscentring of clusters and by uncertainties of the
mass-observable relation (normalization and scatter). We show that available
data (\plk\ maps and the MaxBCG catalogue) should deliver a $7.7\sigma$
detection of the kSZ. A similar cluster catalogue with broader sky coverage
should increase the detection significance to $\sim 13\sigma$. We point out that
such measurements could be binned in order to study the properties of the cosmic
gas and velocity fields, or combined into a single measurement to constrain
cosmological parameters or deviations of the law of gravity from General
Relativity.
\end{abstract}



\section{Introduction}

The cosmic microwave background (CMB) radiation has a prime role in
modern cosmology. Its study not only gives us access to early-Universe
physics and to tight constraints on the parameters of the background
cosmological model, but also allows us to explore the properties of
baryons and dark matter (DM) in the low-redshift Universe. The pioneering
exploration of the CMB was carried out by the $\textit{Cosmic Background
Explorer}$ (\cobe) satellite which provided the first detection of 
temperature fluctuations. More recently, the $\textit{Wilkinson Microwave
Anisotropy Probe}$ (\wmap) and \plk\ satellites have provided ever more
detailed and accurate full-sky CMB anisotropy maps, which have even
been able to detect lensing of the CMB photons by the large-scale
structure of the Universe.

The structure in the CMB radiation can be classified into two
types. `Primary anisotropies' are those resulting from physics
before or on the last scattering surface, whereas `secondary
anisotropies' are those caused by the interaction of CMB photons with
intervening structures at lower redshift. Among the latter, the
Sunyaev-Zel'dovich effects
\citep[SZ;][]{1972CoASP...4..173S,1980MNRAS.190..413S,
  1980ARAA..18..537S} are particularly important and interesting.

The SZ effects refer to the inverse Compton scattering of CMB photons
by free electrons in the hot intracluster and intergalactic gas that
they encounter on their journey from $z\sim1100$ to $z=0$.  This
scattering results in a net energy gain of CMB photons at fixed number
density and consequently distorts their spectrum. This effect is known
as the thermal SZ effect (hereafter tSZ). Motions of the plasma with
respect to the CMB rest frame produces Doppler effects which shift the
temperature of the CMB spectrum while maintaining its blackbody
form. This is known as the kinematic SZ effect (hereafter kSZ). The
tSZ and kSZ imprint characteristic patterns in the CMB sky, which
reflect the structure of the intergalactic gas at (relatively) low
redshifts, so by identifying these patters, we can learn about the
distribution of the baryons at the corresponding epochs.

The tSZ effect provides a measurement of the integral of the electron
pressure along each line-of-sight to the recombination surface.  The
signals detected so far have primarily been due to the hot and dense
gas in the intracluster medium of intervening galaxy groups and
clusters. The kSZ, on the other hand, offers a unique opportunity to
characterize the cosmic peculiar velocity field in the distant
Universe, and to search for the so-called `missing baryons', the
bulk of the cosmic baryon density which apparently lies outside
galaxies and galaxy clusters, and has yet to be identified directly at
$z<2$. Thus, SZ measurements can shed light on a number of important
aspects of the non-linear galaxy formation and feedback processes which
structure the low-redshift Universe, as well giving access to the
cosmic flow field which is influenced by the nature of Dark Energy and
by possible modifications of the theory of gravity
\citep[e.g.][]{2013ApJ...765L..32K}.

SZ measurements are challenging since the signals are small and are
buried beneath primary CMB fluctuations, instrument noise, and
foreground contaminants (e.g.  galactic dust and synchrotron emission,
free--free emission, etc.). The measurement of kSZ effects is
particularly tough, since for galaxy clusters, which provide the
strongest individual signals, the kSZ amplitude is an order of
magnitude smaller than the tSZ, and two orders of magnitude smaller
than primary CMB fluctuations. In addition to this, unlike the tSZ
which has a distinctive spectral signature, the frequency dependence
of the kSZ signal is identical to that of the primary CMB
fluctuations. Furthermore, for an ensemble of clusters, the signal is
predicted to be symmetrically distributed about zero, making it
impossible to enhance the signal to noise by stacking, as is often
done for the tSZ.

Despite these difficulties, the latest generation of CMB telescopes --
the South Pole Telescope \citep{2011PASP..123..568C,
  2011ApJ...743...90S}, the Atacama Cosmology Telescope
\citep{2007ApOpt..46.3444F, 2011ApJS..194...41S}, and the
\plk\ satellite -- have achieved high-resolution measurements of the
CMB at millimetre wavelengths over large areas, which is enabling
detailed studies of the SZ effect. In particular, the tSZ has
been detected at high significance and is currently posing interesting
challenges to our current understanding of structure formation and
cosmological parameters \citep{2013arXiv1303.5089P,
  2013arXiv1303.5081P}.

There have also been several claims of detection of the kSZ. Samples
of (X-ray) detected clusters combined with \wmap\ CMB maps have been
used to estimate cosmic bulk flows in \citet{2010ApJ...712L..81K,
  2011ApJ...732....1K, 2011ApJ...737...98O, 2011ApJ...736..116M,
  2012ApJ...758....4M}. These results appear to be in tension with 
$\Lambda$ cold dark matter ($\Lambda$CDM), but the discrepancy may not be as severe as claimed
\citep{2012ApJ...761..151L}, since it has not been confirmed by new
\plk\ results \citep{2014A&A...561A..97P}.  More recently, a
$3.8\sigma$ kSZ detection has been reported from correlations of CMB
residuals about pairs of luminous red galaxies
\citep{2012PhRvL.109d1101H}. This finding is in qualitative agreement
with the $\Lambda$CDM expectations as inferred from cosmological
hydrodnamics simulations \citep{2013MNRAS.432.1600D}. Another approach
is to use linear perturbation theory to estimate the cosmic flow field 
from a three-dimensional distribution of galaxies, and in this way
obtain a template for the expected kSZ signal on the sky
\citep{2009arXiv0903.2845H, 2011MNRAS.413..628S}. There has been a
marginal detection of the kSZ from applying this method to the 2MASS
survey \citep{2013MNRAS.430.1617L}. All these examples illustrate the
potential of the field and show that the quality of the data is
reaching a level where cosmological and astrophysical exploitation of
the kSZ effect is imminent.

In this paper, we develop and test a new but related statistical
method to measure the kSZ signal. The idea is to combine a sample of
independently detected galaxy clusters with a velocity field estimated
by applying perturbation theory to the galaxy distribution. These two
ingredients allow construction of a matched filter that optimally
weights the signal from each cluster based on the noise in the CMB and
velocity maps and the signal amplitude predicted from cluster scaling
relations and the velocity reconstruction itself.  We investigate the
various sources of uncertainty in this measurement and show that our
approach should yield a kSZ detection with high statistical
significance ($7.7\sigma$), even with current data sets. An advantage
of this scheme with respect to previous ones is that it allows kSZ
measurements to be grouped into different mass bins to study the gas
properties of galaxy clusters. Alternatively, they can be combined
into a single measurement to constrain the relation between density
and velocity fields, giving information about the law of gravity and
about cosmological parameters.

Our paper is organized as follows. We first present our
statistical methods, including the derivation of the matched
filter for kSZ measurements (\S2). In \S3, we describe the way in which
we create mock CMB skies including the kSZ effects expected for a
realistic sample of clusters. We provide details of the application of
our approach to mock data in \S4. In \S5, we present our results, and
explore and quantify different sources of systematic errors. In the
final section, \S6, we discuss our results and conclude.

\section{Optimal measurement of the kSZ effect}
\label{s_tool}

In this section, we will present and discuss our method to measure the
kSZ effect for a given set of galaxy clusters. 

\subsection{Matched Filter}
\label{mfilter}

As mentioned before, the typical amplitude of the kSZ effect is smaller than
the tSZ and than the primordial CMB temperature fluctuations. Thus, it is
necessary to develop the best possible estimator of the signal given all the
sources of noise. Here, we choose to follow the so-called matched filter
formalism. 

A matched filter is a linear processing of the data, specifically designed to
maximize the signal-to-noise ratio (SNR) for a set of known template signal and
(additive and stochastic) noise power spectra. For the case we consider here,
this means to optimally extract the kSZ signal from clusters assuming the
expected signal profile, the power spectrum of CMB fluctuations, and the
uncertainties in the estimates for the velocity and mass of clusters.

The first step in the formalism is to define a signal template. This is simply
the expected kSZ signal, whose amplitude and spatial distribution for a galaxy
cluster are given by

\begin{eqnarray}
 \label{eq:eqn_ksz}
 \left(\frac{\Delta T}{T_{\mathrm{CMB}}}\right)_{\rm{kSZ}}(\bmath{\theta}) & \equiv & k(\bmath{\theta}) \nonumber \\
 & = & -\frac{\sigma_{\mathrm{T}}}{c} \int a \, \dd \chi \, n_\e (\bmath{\theta},\chi) \, v_\mathrm{r}(\bmath{\theta},\chi) \mbox{ .}
\end{eqnarray}

Here, $\sigma_{\mathrm{T}}$ is the Thomson cross-section and $c$ is the speed of
light, $a$ is the expansion factor, $\chi$ is a line-of-sight distance in
comoving coordinate, $v_\mathrm{r}$ represents the velocity of the gas along
the line of sight, and $n_\mathrm{e}$ is the number of free electrons both as a function
of $\theta$, the angular position on the sky. The minus sign follows the
convention that CMB photons gain energy when the free electrons move towards
us, and thus the temperature of CMB photons increases. 

Assuming that (i) the spatial distribution on the sky of free electrons inside a
cluster can be described as a projected NFW profile, (ii) that the velocity
field has a large correlation length (much larger than the extent of a
cluster), and (iii) that the gas is fully ionized, we obtain:

\begin{eqnarray}
 \label{eq:eqn_ksz_c}
 k(\theta) = -\frac{\sigma_{\mathrm{T}} v_c}{c}\, f_{\rm b}\, \mu \, \Sigma_{\mathrm{NFW}}(\theta) ,
\end{eqnarray}

\noindent where $v_c$ is the line-of-sight velocity of the cluster, $f_\mathrm{b}$ is the
(cosmic) baryon fraction, $\mu$ is the number of electrons per unit of gas mass.
For building the filter, we choose the spatial template profile to be

\begin{equation}
\label{eq:profile}
  \tau(x) = \frac{A}{(cx)^2-1}\begin{cases}
    1-{2 \over \sqrt{1-(cx)^2}}\ \mbox{tanh}^{-1} \sqrt{{1-cx\over cx+1}} & 0<x<1 \cr
    0 & x=1 \cr
    1-{2 \over \sqrt{(cx)^2-1}}\ \mbox{tan}^{-1} \sqrt{{cx-1\over cx+1}} & x>1 ,
  \end{cases}
\end{equation}

\noindent $c$ is the cluster's concentration parameter, $x =
r/r_{200}=\theta/\theta_{200}$ the dimensionless radius. $A$ is a constant
normalizing the template profile at $x=0$,  so when this filter is located on
the centre of a cluster, it will return a statistically unbiased amplitude of
the kSZ signal. We note that our approach and results do not depend on assuming
this particular functional form, the only requirement is to the correct profile
to be know. Our choice (a projected NFW profile) is justified here since in our
forthcoming tests we assume that the spatial distribution of baryons follows
that of the DM. However, when applied to real data, a different,
observationally motivated profile might be preferred.

\begin{figure}
\plotone{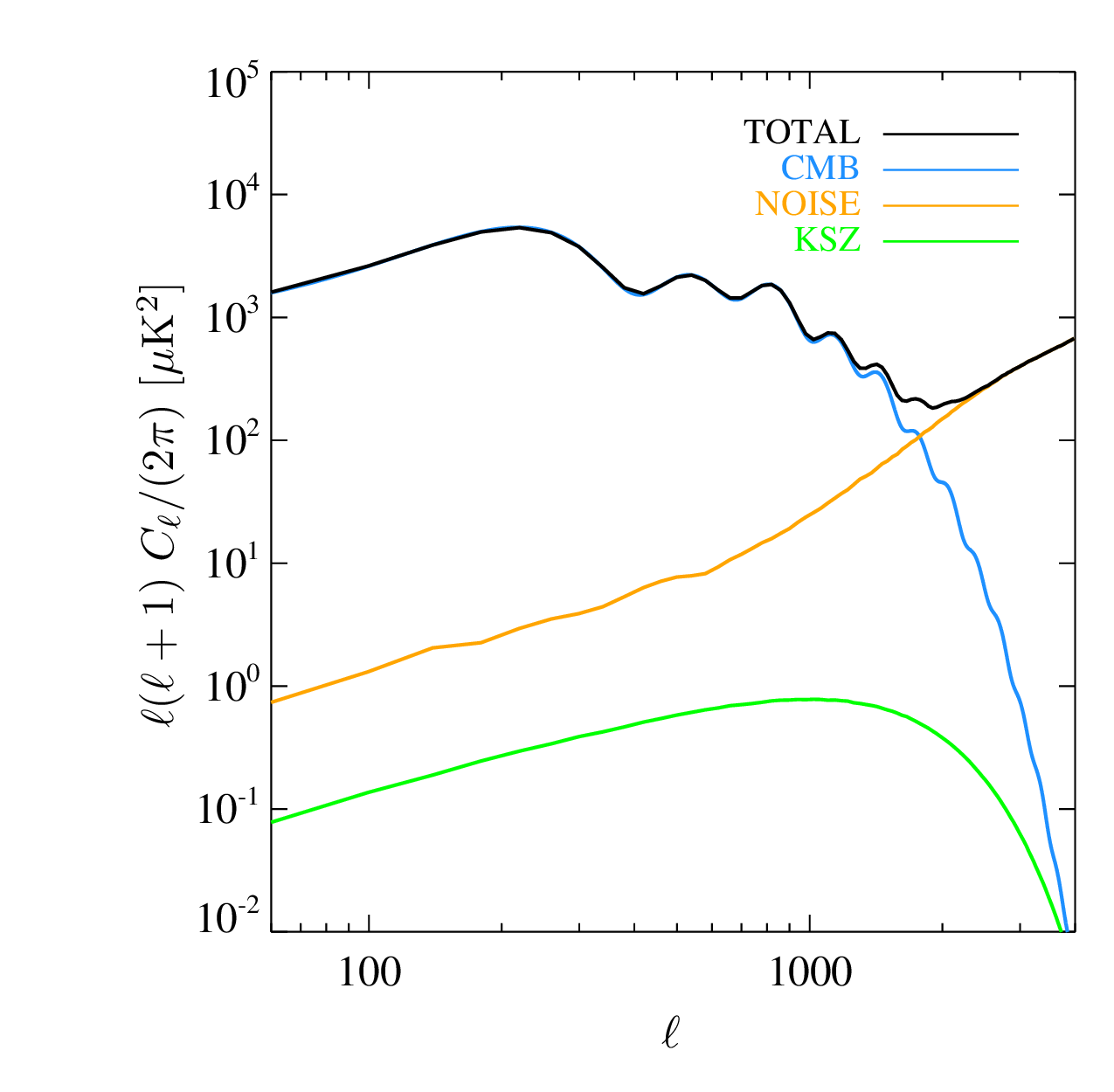}
\caption{Angular power spectra of one realization of our simulated \plk\
SMICA-like sky maps. Three components are shown as CMB (blue solid line),
instrument noise (orange solid line) and KSZ (green solid line). All power
spectra have been convolved with a beam function with FWHM$=$5 acrmin.}
\label{map_spectra}
\end{figure}

The next step is to define $P(\bmath{k})$, the power spectrum of the noise. In
Fig.~\ref{map_spectra}, we show the contribution to the total angular CMB power
spectrum of different components for a \plk-like experiment: primordial
anisotropies (blue line), instrumental noise (orange line) and kSZ (green
line). Here we can see that the kSZ signal is sub-dominant at all scales. In
consequence, we approximate the noise in our kSZ estimates as the power
spectrum of primordial CMB fluctuations plus the noise contribution, that is
$P=P_{\mathrm{CMB}}|\hat{B}|^2 + P_{\mathrm{noise}}$. 

The shape of the matched filter is set by requiring a minimum variance
estimator. Following \citet{1996MNRAS.279..545H, 2005A&A...429..417M,
2006A&A...459..341M}, in our case it is possible to show that the Fourier
transform of the filter is given by

\begin{equation}
\label{eq:filter}
{\hat{\Psi}}({\vect{k}}) = \sigma^2 \frac{\hat{\tau} (\vect{k}) \hat{B} (\vect{k})}{P(\vect{k})},
\end{equation}

\noindent where $\hat{\tau}$ is the Fourier transform of the signal profile,
$\hat{B} (\vect{k})$ is the beam function of a given CMB experiment which we
assume follows a Gaussian profile. The variance of the filtered input data 
is denoted by $\sigma^2$: 

\begin{equation}
\label{eq:filternoise}
\sigma^2 = \left [ \int \frac{{|\hat{\tau}(\vect{k})\hat{B}(\vect{k})|}^2}{P(\vect{k})} \frac{\dd^2k}{(2 \pi)^2} \right ]^{-1}.
\end{equation}

\begin{figure}
\plotone{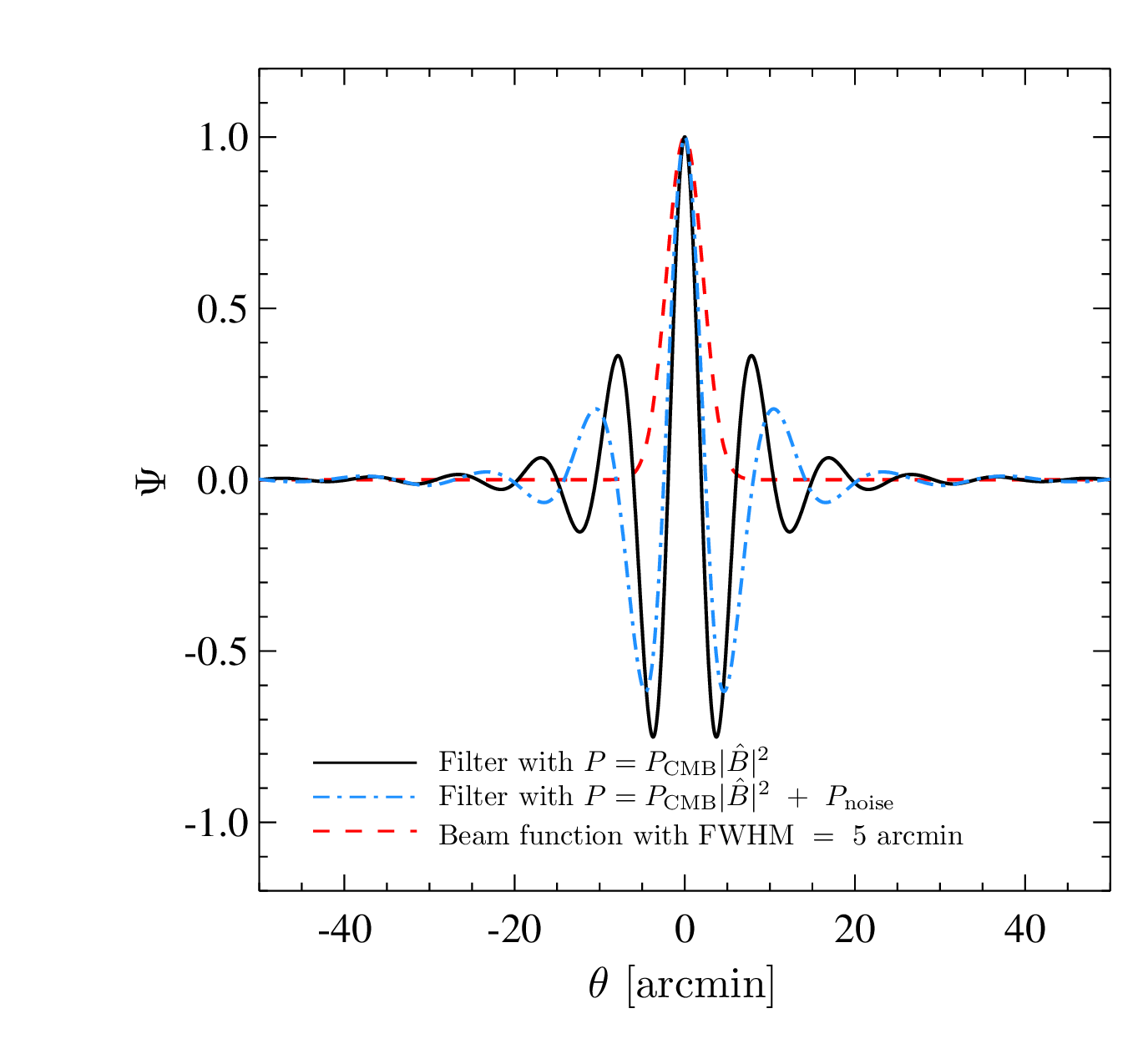}
\caption{An example of matched filter for a cluster with a angular scale
$\theta_{200}= 10$ arcmin. The black solid line shows the radial filter profile
when noise power spectrum $P(\vect{k})$ in equation~(\ref{eq:filter}) only contains CMB
component. The filter when both CMB and instrument noise are contained in
$P(\vect{k})$ is shown as blue dotted$-$dashed line. The red dashed line gives a
reference of a beam function with FWHM$=$5 acrmin.}
\label{fprof}
\end{figure}

In Fig.~\ref{fprof}, we show the resulting filter $\Psi(\theta)$, for a cluster
(with mass around $10^{14}\msun$ and $z\sim0.1$) with an angular size of 10
arcmin on the sky. By comparing the solid black and dot$-$dashed blue lines, we
can see how the filter is modified when the instrument noise in considered in
addition to the primary anisotropies in the CMB. For comparison, we show the
assumed beam profile as a dashed red line. We note that we have checked the impact
of uncertainties in the determination of the centre of clusters (c.f.
\S\ref{miscentering}). In this case, the amplitude of the filter changes, but
its shape remains largely unaffected due to the dominating effect of the beam
size of the \plk\ experiment. This, however, might be different for
higher resolution CMB experiments.

In general, the central value returned by the filter refers to the signal
integrated over a patch on the sky of a given radius (a cone in three
dimensions). Here, we choose to integrate up to three times the size of the
target cluster, though our results are not sensitive to the exact integration
limit. Due to the large coherence length of the cosmic velocity field,
integrating outside the clusters boundary has the advantage of including
material that is likely to be moving with the target cluster (c.f. \S3.2).
Thus, our kSZ measurement corresponds to $K^{\rm{cyl}}_{3r_{200}}$, the total
signal within a cylinder of aperture radius $3\times r_{200}$ (where $r_{200}$
is the radius containing a mean density equal to 200 times the critical value
in the Universe). This measurement can be scaled to the expected signal
produced by a spherical halo, $K_{200}$, by the following quantity: 

\begin{equation}
  \frac{K^{\rm{cyl}}_{3r_{200}}}{K_{200}} = \frac{\int_0^\infty \dd r \int_{r \sin \theta < 3r_{200}} \dd \theta \, \rho(r) 2 \pi r^2}{\int_0^{r_{200}} \dd r \rho(r) 4 \pi r^2},
\end{equation}

\noindent where $\rho(r)$ is given by $\rho(x)=\frac{\rho_0}{x(1+x)^2}$,
$\rho_0$ is a characteristic density, and $x = r/r_{200}$. Supposing cluster mass
$M_{200}$ (cluster mass within $r_{200}$) can be inferred from other observed
properties, we can obtain an estimation of cluster's velocity through relation

\begin{eqnarray}
 \label{eq:eqn_ksz_c}
 K_{200} = -\frac{\sigma_{\mathrm{T}} v_c}{c}\, f_{\rm b}\, \mu \, M_{200}.
\end{eqnarray}

At this point, this estimator also contains contributions from the tSZ
effect, point sources, etc. As we will see next, if there is an external
estimate for the velocity field, then these extra terms will vanish and we can
recover a clean measurement of the kSZ effect which can be used to constrain
the law of gravity, cosmological parameters and gas properties in our cluster
catalogue.

\subsection{The velocity field}

In linear theory, and assuming a linear bias, the peculiar velocity field is
directly proportional to the logarithmic derivative of the growth rate and to
the galaxy overdensity field. Explicitly, the Fourier transform of the velocity
field is

\begin{equation}
\label{eq:vel_lin}
\vect{v}(\vect{k}) = -\vect{i} \beta(z) H_0 \delta_\mathrm{g}(\vect{k}) \frac{\vect{k}}{k^2}~,
\end{equation}

\noindent with $\beta(z) = f(\Omega_\mathrm{m},z)/b_\mathrm{g}(z)$, $b_\mathrm{g}$ is the galaxy bias, $H_0$ is
the Hubble constant and $f(\Omega_\mathrm{m},z) \equiv \frac{\dd\ln D(z)}{\dd\ln a}$, where 
$D(z)$ is the growth factor and $a$ is the expansion factor.

Thus, on large scales (where these relations hold) the observed galaxy
distribution can be used to obtain an estimate of the velocity of galaxy
clusters. Note also that there are equivalent expressions in higher order
perturbation theory. These can achieve higher accuracies and are valid down to
smaller scales \citep{2012MNRAS.425.2422K}.

Contrasting the reconstructed velocities with the measurements obtained by our
matched filter for a cluster $i$, one can constrain a parameter $\alpha$:

\begin{eqnarray}
\label{eq:beta_est_i}
\alpha_i &=& - \frac{c}{\sigma_T\,f_\mathrm{b}\,\mu} \frac{K_{\mathrm{200},i}}{M_{200}}\frac{1}{v_{\mathrm{rec},i}}, \\
         &=& \frac{v_{\rm{kSZ},i}}{v_{\rm{rec},i}} + \frac{\epsilon_i}{v_{\rm{rec},i}},
\end{eqnarray}

\noindent where $\epsilon$ captures all other sources to CMB temperature
fluctuations inside clusters (e.g. tSZ, point sources, etc.). The measurement
from individual clusters can be combined into a single measurement of the
$\alpha$ parameter

\begin{equation}
\label{eq:beta_est}
    \alpha =  \frac{\sum_i \alpha_i \, w_i}{\sum_i w_i},
\end{equation}

\noindent with the associated error

\begin{equation}
\label{eq:beta_error}
 \sigma_{\alpha} = \left [ \frac{1}{\sum_i w_i} \right ]^{1/2}.
\end{equation}

Since $\epsilon$ is expected to be uncorrelated with the velocity field, and the
sign of $v_{\rm{rec}}$ changes from one cluster to another, so the expectation
value of $\epsilon/v_{\rm{rec}}$ is zero. Therefore if $\alpha$ is equal to the
unity, this means that the gravity model and cosmological parameters assumed are
supported by the kSZ data. Otherwise, $\alpha \neq 1$, a different model is
preferred. In other words, this ratio constrains directly $\beta/\beta_{\rm
fid}$, where $\beta_{\rm fid}$ is the fiducial value of $\beta$ assumed in
computing the velocities from the galaxy distribution. 

It is important to emphasize that this method is based on the assumption that
reconstructed velocity has the correct sign, so that after effectively
weighting individual kSZ measurements by the SNR
expected for each cluster, all available signals are optimally combined
together. For instance, regions with a velocity close to zero, are expected to
contribute mostly to the noise, not the signal, and are therefore given less
importance in the final measurement.

Additionally, the weight factors can be modified to include all uncertainties
affecting the measurement. In our case, there are two major sources: one is
intrinsic to the kSZ measurement (we label it with $\sigma_{\rm{kSZ}}$), the
second is in the uncertainty in the velocity estimation due to a given
reconstruction method (labelled as $\sigma_{\rm{rec}}$). If the two
contributions are uncorrelated, the weight assigned to each cluster is

\begin{equation}
\label{eq:weight1}
  w^{-1}_i = \left(\frac{1}{v_{\rm{rec},i}}\right)^2 \left( \sigma^2_{\rm{kSZ},i} 
           + \sigma_{\epsilon}^2 
           + \beta^2_{\rm{fid}} \sigma^2_{\rm{rec},i} \right).
\end{equation}

Both of $\sigma_{\rm{kSZ}}$ and $\sigma_{\rm{rec}}$ vary from cluster to
cluster, and $\sigma_{\epsilon}^2$ is approximately proportional to the inverse
of the number of systems averaged over. If we assume that the uncertainty
arises mainly due to the kSZ measurement, then equation~(\ref{eq:weight1}) is simplified
to

\begin{equation}
\label{eq:weight2}
    w^{-1}_i = \left(\frac{1}{v_{\rm{rec},i}}\right)^2 \sigma^2_{\rm{kSZ},i}.
\end{equation}

For the remainder of this paper, we will present our results based on this
simplified form of the weight factor. We further discuss uncertainty related to
velocity reconstruction in \S\ref{vel_uncer}.

\section{Mock observations}
\label{data}

In this section, we describe the mock kSZ observation we have created to test
and assess the performance of our approach.

\subsection{The {\mxxl} simulation}
\label{mxxl}

We build kSZ mocks based on the Millennium-XXL ({\mxxl}) simulation
\citep{2012MNRAS.426.2046A}. The {\mxxl} simulation uses $6720^3$ particles to
follow the distribution and evolution of DM within a cubic volume with
a comoving side length of $3 \lgpc$. The mass of each simulation particle is
$m_{\rm{p}}=6.17 \times 10^9 \msun$, thus we resolve galaxy clusters with tens
of thousands of particles. The cosmological parameters match those of the
previous two Millennium simulations \citep[]{2005Natur.435..629S,
2009MNRAS.398.1150B}. 

The {\mxxl} simulation combines a large volume and a relatively high-mass
resolution.  Simultaneously fulfilling these conditions posed a serious
challenge to supercomputational facilities in terms of raw execution time, RAM
requirements, I/O load and long-term storage. In order to alleviate these, the
full particle data were stored at redshifts $z = 0$, $0.25$, $1$ and $3$.
Self-bound halo and subhalo catalogues, among other data products were produced
on-the-fly. For further information, we refer the reader to
\citet{2012MNRAS.426.2046A}.

Most important for our purposes is the fact that this simulation produces a
suitable DM backbone for our kSZ modelling. This provides a realistic
catalogue of DM clusters as well as a fully non-linear velocity field
with all the features and correlations we expect in $\Lambda$CDM, including the
non-negligible contribution of large Fourier modes.

\subsection{Light-cone and the kSZ effect}
\label{ksz}

We build a light-cone using the $z=0.25$ snapshot from the {\mxxl}, and
considering all particles within a sphere of $1500 \lmpc$ radius. This produces
an all-sky light-cone up to $z=0.56$, without any repetition of the simulation
box. Note that this procedure effectively neglects the evolution of the mass
clustering along the line of sight; however, this is a reasonable approximation
given the restricted redshift range we consider. We also build a light-cone
with the position and velocities of all haloes in our catalogue.
  
Then, we assume that all the gas in the Universe is ionized and that the
position and velocity of baryons follow those of DM. This is a
reasonable approximation on large and intermediate scales \citep{Angulo2013}.
Thus, the kSZ effect integrated over a area element in our simulated sky is
given by the discrete version of equation~(\ref{eq:eqn_ksz})

\begin{equation}
 \label{eq:eqn_ksz_d}
 \left(\frac{\Delta T}{T}\right)_{\rm{kSZ}}=k=-\frac{\sigma_{\mathrm{T}}\,f_\mathrm{b}\,\mu}{c}\sum_{i}\frac{v_{\mathrm{r},i}\ m_{\mathrm{dm},i}}{\dd\Omega_{\mathrm{pix}} D_{\mathrm{a},i}^2}\ \mbox{ ,}
\end{equation}

\noindent where the summation runs over all particles that contribute to the 
given area element on the sky, $D_{\mathrm{a}}$ is the angular diameter distance
and $\dd\Omega_{\mathrm{pix}}$ is the solid angle of the area element.

We pixelize our sky map using the $\healpix$ software
\citep{2005ApJ...622..759G}\footnote{{\tt http://healpix.jpl.nasa.gov/}} with
$N_{\rm{side}}=2048$ pixels. This corresponds in total to 50 331 648
elements, each of which covers an area equal to $1.43^{-5}\deg^2$.

\begin{figure*}
\plotsiderotate{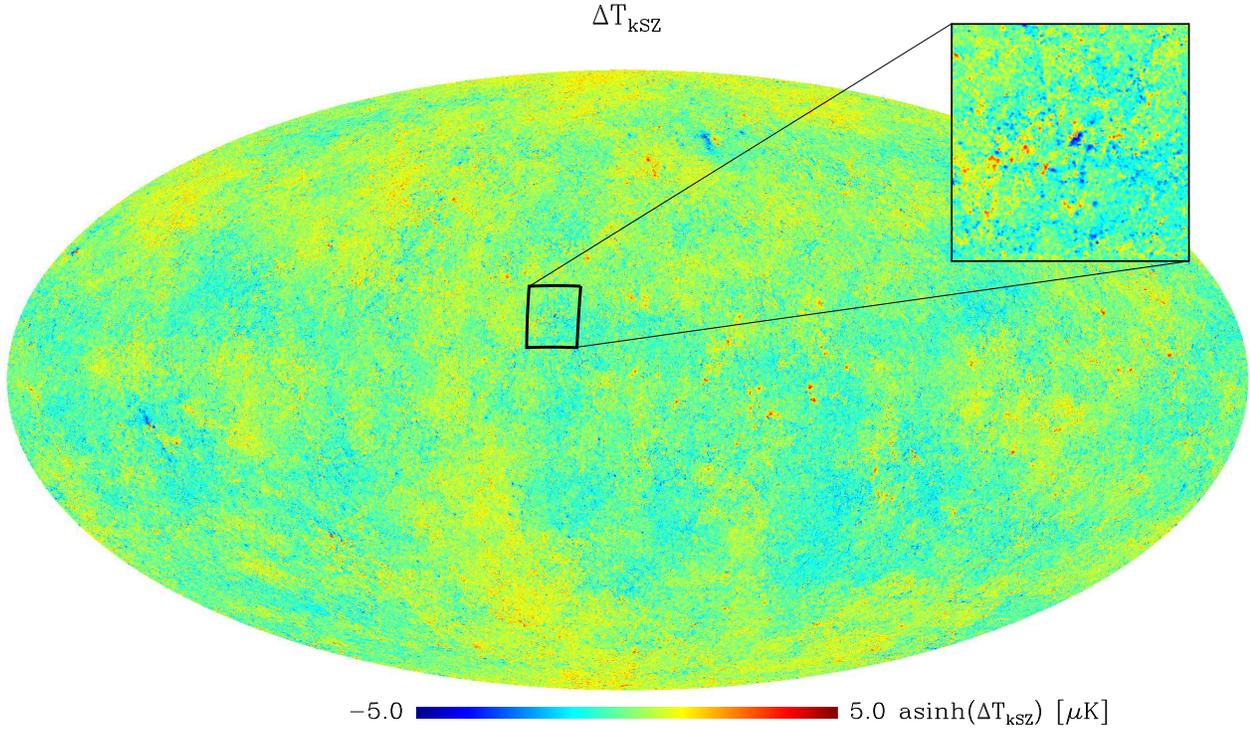}
\caption{Simulated all-sky map of the kSZ signal with a resolution of
$N_\mathrm{side}=2048$. The map is smoothed with a beam function with FWHM$=$5
acrmin and is colour-coded by arcsinh($\Delta T_\mathrm{kSZ}$). The overlaid
panel shows a patch with side length of $14^{\circ}.66$, zooming in around a
prominent structure which produces a clear kSZ signal.}
\label{kszmap}
\end{figure*}

In Fig.~\ref{kszmap} we show a Mollweide representation of our kSZ sky. The
mean of the map corresponds to a value of $\langle \Delta T_\mathrm{kSZ}
\rangle=0.12\ \mu \rm{K}$ with variance $\sigma=1.36\ \mu \rm{K}$. The actual
power spectrum of the simulated kSZ is shown by the green line in
Fig.~\ref{map_spectra}. Note that the map shows a large coherence length,
with regions of similar amplitude extending over large fractions of the sky.
This is a consequence of the large correlation length of the velocity field
expected in CDM density power spectra, where velocity fluctuations receive
significant contributions from very large modes. The inset in this figure 
shows a zoom to the fluctuations inside a $16\deg.6$ patch (approximately
$200\,\rm{Mpc}$ wide at $z=0.25$). 

Finally, we mimic \plk\ CMB observations. The \plk\ satellite observes the sky
in nine frequency channels from 30 to 875 GHz, the angular resolution ranges
from 33 arcmin for the lowest frequency channel down to 5 arcmin for the
highest. In this work, we only focus on the SMICA-like CMB map, which is a
foreground-cleaned map. In consequence, we directly simulate the sky map that only
contains CMB and kSZ components. After generating the
map, we smooth the map with a Gaussian kernel with FWHM$=$5 arcmin (angular
resolution of \plk\ SMICA map). Posteriorly, we will also include the
corresponding instrument noise.

\subsection{Cluster Catalogue}
\label{cluster}
\begin{figure}
\plotone{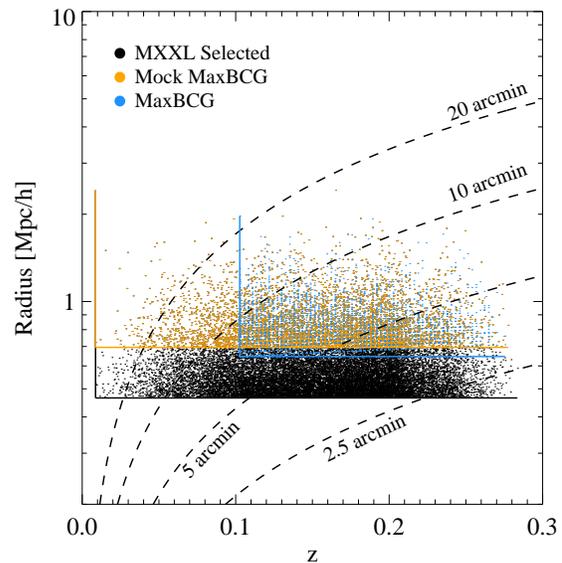}
\caption{The distribution of clusters used in this work on the radii and
redshift plane. The black dots and yellow dots are clusters in each 
catalogue. The MaxBCG clusters are also shown (blue dots), assuming the same
cosmology as {\mxxl} simulation. The solid lines around clusters  indicate selection
boundaries we use to construct the catalogues. The black dashed lines are here
to lay out the relation of cluster radius for fixed angular size as a function
of redshift.}
\label{halo_cata}
\end{figure}

For our analysis, we consider two different cluster catalogues. We restrict 
the samples to a volume similar to that of the SDSS, over which there are
reconstructed density and velocity fields \citep{2009MNRAS.400..183K,
2010MNRAS.409..355J}.

\begin{itemize}
\item[1)] MXXL selected: our first sample contains all haloes in our light-cone
with mass above $1.5\times10^{13}\msun$. This contains $24529$ objects. 

\item[2)] Mock MaxBCG: our second sample employs a higher mass cut,
$5\times10^{13}\msun$, which roughly corresponds to a threshold in optical
richness of $10$ for the MaxBCG catalogue \citep{2007ApJ...660..239K}. This
extra condition reduces the number of clusters in this catalogues down to
$5663$. 
\end{itemize}

We summarize the main properties of our samples in Table~\ref{mtab}. In
Fig.~\ref{halo_cata}, we show the redshift and size of the clusters in our
samples. These properties will help us to understand the contribution of
different types of clusters to the total SNR for the kSZ measurements.  We can
see that given our selection criteria most systems are found at redshifts below
0.3. This validates the redshift range covered by our kSZ light-cone.  For
comparison, we also show the properties of MaxBCG cluster catalogue.  We note
that the mass cut-off of our Mock MaxBCG catalogue roughly coincides with the
observational catalogue. Because no selection function is applied, our Mock-
MaxBCG sample contains 28 per cent more clusters than the real one, most of these
additional clusters have redshifts below $\sim0.1$. 

Another issue should be noted is that our catalogues are built from a single
snapshot at $z=0.25$, the mass cut shows a constant radius threshold across
whole redshift range in Fig.~\ref{halo_cata}. But cluster with the same mass
does not have a constant physical radius at different redshift, because the
$M_{200}$ and $R_{200}$ are linked by $\rho_{\rm{crit}}(z)$. So for real
cluster catalogue, the mass cut-off should not correspond to a constant radius
threshold. But this is a minor issue, and will not affect our analysis and
results.

\subsection{Reconstructed velocity field}

As discussed before, the peculiar velocity of galaxy clusters can be estimated
using perturbation theory and a three-dimensional distribution of galaxies.
Naturally, there are uncertainties associated with this procedure, thus, in order
to explore the impact of these, for our analysis, we consider three different
types of velocity fields.

\begin{enumerate}

\item[1)] $v_{\rm{halo}}$: These correspond to the true velocity of the cluster,
as computed by the centre of mass velocity of the parent FOF halo. Naturally,
this corresponds to the best possible estimation, and it is useful to
differentiate the impact of the uncertainties in the velocities from other
sources.

\item[2)] $v_{\rm{rec}}$: These correspond to velocities estimated from the DM 
density field with linear perturbation theory. In practice, we compute
these by the mapping DM particles on to a grid using a Clouds-in-Cell (CIC)
assignment scheme, with a spatial resolution of $1.5\lmpc$. Then, we smooth the
density field with a Gaussian kernel of size $r_\mathrm{s}=[2.5, 5, 10]\lmpc$,
and use the smoothed field as a source in equation~(\ref{eq:vel_lin}) to obtain an
estimate for the velocity field. Finally, we interpolate this field to the
positions of clusters.

\item[3)] {\bf $v_{\rm{CIC}}$}: We generate another velocity field by directly
mapping the velocity of DM particles on to a grid using the CIC
assignment scheme. We then smooth this field using Gaussian kernels of size
$r_\mathrm{s}=[2.5, 5, 10]\lmpc$, and interpolate back to the clusters
positions.

\end{enumerate}

\begin{table}
  \caption{Halo catalogue used in this work}
  \label{mtab}
  \begin{tabular}{@{}lcccc}
   \hline
             & $M_{200}$ range \vspace{-0.1cm} & $\theta_{200}$ range \vspace{-0.1cm} &        \\
   Cata. name & \vspace{-0.1cm}                 & \vspace{-0.1cm}                      & Number \\
             & $(\times 10^{10} \msun)$        & (degree)                             &        \\
   \hline
   \ MXXL selected    & (1500, 210885) & (0.035, 1.897) & 24529 \\
   \ Mock MaxBCG    & (5000, 210885) & (0.052, 1.897) & 5263  \\
   \hline
   \ MaxBCG         & (5082, 144073) & (0.052, 0.374) & 4058  \\
   \hline
\end{tabular}

\medskip
Note: For clusters in real MaxBCG catalogue, the cluster mass is computed with
$M_{200}-N_{200}$ provided by \citet{2010MNRAS.404..486H}. The cluster angular
size is computed under the same cosmology model as {\mxxl} simulation.
\end{table}
 
\section{kSZ and $\alpha$ Measurements}

Now we are in position to apply our matched filter procedure to the simulated
kSZ+CMB sky, and using the different cluster catalogues and velocity estimates
discussed in the previous section.

At the position of each cluster, we convolve the CMB maps with the matched
filter. We do this in Fourier space and on a patch of side length of
$14^{\circ}.66$ and $512\times512$ pixels, which makes patches have the same
angular resolution as the original sky map. The patch size is chosen to be
large enough to ensure a representative assessment of background noise. The
characteristic size of the filter is set by the cluster's apparent size on the
sky and by using the concentration$-$mass relation of  proposed by
\citet{2008MNRAS.390L..64D}. 

The value of $\alpha$ is estimated for each cluster in our catalogues
(equation~(\ref{eq:beta_est_i})) and the combined measurement is given by taking the
ensemble weighted mean (equation~(\ref{eq:beta_est})).

In order to assess the advantages of the matched filter approach, we perform
another measurement of the kSZ signal using a simple aperture photometry method
(AP filter). The kSZ flux is estimated as the total kSZ flux within a circular
area of radius $R_1$ minus the expected background, which is set by the average
kSZ flux in a annulus of dimensions $R_1=3\,\times r_{200}$ and
$R_2=\sqrt{2}\,R_1$.  Therefore, the kSZ signal within a cylinder of aperture
radius $3\times r_{200}$, (analogous to the quantity measured by the matched filter)
is given by

\begin{equation}
  K^{\rm{cyl}}_{3r_{200}} =  K_{[0,R_1]} - K_{[R_1,R_2]}.
\end{equation}

Following \citet{2014A&A...561A..97P}, the uncertainty for the measurement
about each cluster is set as the rms fluctuation of the AP filter applied at
100 randomly chosen positions. As in the case of matched filters, the value
of $\alpha_i$ for each cluster is the weighted by its uncertainty and in this
way a global $\alpha$ is computed.

\section{Results}
\label{results}

\begin{table*}
\begin{minipage}{0.95\textwidth}
  \caption{Estimated mean $\alpha$ value comparison with 50 realizations of sky, between cluster catalogues, Ap filter and matched filter, with/without instrument noise}
  \label{tab_result1}
  \begin{tabular}{@{}l|cccc|cc}
  \toprule
                                       & \multicolumn{4}{c}{Mock MaxBCG} & \multicolumn{2}{c}{MXXL selected}\\
   \cmidrule(lr){2-5}
   \cmidrule(lr){6-7}
                                       & \multicolumn{6}{c}{\vspace{-0.1cm}}\\
   $v$ used in reference               & CMB          & CMB + Noise  & CMB & CMB + Noise & CMB & CMB + Noise \\
                                       & AP filter & AP filter &           &             &     &             \\
   \midrule
   $v_{\rm{halo}}                    $ & $1.028 \pm 1.103$ & $1.026 \pm 1.106$ & $0.963 \pm 0.046$ & $1.011 \pm 0.096$ & $0.977 \pm 0.035$ & $1.029 \pm 0.075$ \\
   \midrule                                                                                                                                   
   $v_{\rm{rec},r_{\rm{s}}=2.5 \lmpc}$ & $0.792 \pm 0.933$ & $0.788 \pm 0.935$ & $0.794 \pm 0.038$ & $0.837 \pm 0.080$ & $0.831 \pm 0.030$ & $0.875 \pm 0.065$ \\
   $v_{\rm{rec},r_{\rm{s}}=  5 \lmpc}$ & $0.967 \pm 1.131$ & $0.964 \pm 1.134$ & $0.969 \pm 0.046$ & $1.018 \pm 0.097$ & $0.988 \pm 0.036$ & $1.040 \pm 0.078$ \\
   $v_{\rm{rec},r_{\rm{s}}= 10 \lmpc}$ & $1.040 \pm 1.352$ & $1.041 \pm 1.356$ & $1.072 \pm 0.055$ & $1.125 \pm 0.115$ & $1.082 \pm 0.043$ & $1.138 \pm 0.092$ \\
   \midrule                                                                                                                                   
   $v_{\rm{CIC},r_{\rm{s}}=2.5 \lmpc}$ & $1.039 \pm 1.163$ & $1.036 \pm 1.166$ & $1.021 \pm 0.048$ & $1.071 \pm 0.100$ & $1.029 \pm 0.037$ & $1.084 \pm 0.080$ \\
   $v_{\rm{CIC},r_{\rm{s}}=  5 \lmpc}$ & $1.055 \pm 1.261$ & $1.054 \pm 1.264$ & $1.051 \pm 0.051$ & $1.101 \pm 0.108$ & $1.058 \pm 0.040$ & $1.114 \pm 0.086$ \\ 
   $v_{\rm{CIC},r_{\rm{s}}= 10 \lmpc}$ & $1.083 \pm 1.428$ & $1.086 \pm 1.432$ & $1.105 \pm 0.058$ & $1.161 \pm 0.121$ & $1.108 \pm 0.045$ & $1.168 \pm 0.097$ \\
  \bottomrule
\end{tabular}

\medskip
\end{minipage}
\end{table*}

\begin{table*}
\begin{minipage}{0.95\textwidth}
  \caption{Estimated mean $\alpha$ value with 50 realizations of considered mass scatter and miscentring effects.}
  \label{tab_result2}
  \begin{tabular}{@{}l|cccccc}
  \toprule
                         & \multicolumn{6}{c}{Mock MaxBCG} \\
   \cmidrule(lr){2-7}
                         & \multicolumn{5}{c}{\vspace{-0.1cm}}\\
   $v$ used in reference               & CMB + Noise & CMB + Noise & CMB + Noise    & CMB + Noise   & CMB + Noise   & CMB + Noise \\
                                       &             & + velocity  & + Mass scatter & + miscentring & + miscentring & + miscentring with correction \\
                                       &             & systematics &                & without correction & with correction   &+ Mass scatter \\
   \midrule
   $v_{\rm{halo}}                    $ & $0.959 \pm 0.096$ & $0.957 \pm 0.097$ & $0.970 \pm 0.099$ & $0.730 \pm 0.097$ & $0.955 \pm 0.126$ & $1.026 \pm 0.130$ \\
   \midrule                                                                  
   $v_{\rm{rec},r_{\rm{s}}=2.5 \lmpc}$ & $0.757 \pm 0.080$ & $0.757 \pm 0.081$ & $0.776 \pm 0.083$ & $0.588 \pm 0.082$ & $0.751 \pm 0.105$ & $0.793 \pm 0.107$ \\
   $v_{\rm{rec},r_{\rm{s}}=  5 \lmpc}$ & $0.951 \pm 0.097$ & $0.949 \pm 0.099$ & $0.971 \pm 0.101$ & $0.734 \pm 0.099$ & $0.952 \pm 0.128$ & $1.014 \pm 0.132$ \\
   $v_{\rm{rec},r_{\rm{s}}= 10 \lmpc}$ & $1.111 \pm 0.115$ & $1.112 \pm 0.117$ & $1.137 \pm 0.119$ & $0.862 \pm 0.117$ & $1.127 \pm 0.152$ & $1.202 \pm 0.156$ \\
   \midrule                                                                  
   $v_{\rm{CIC},r_{\rm{s}}=2.5 \lmpc}$ & $1.014 \pm 0.100$ & $1.011 \pm 0.102$ & $1.029 \pm 0.104$ & $0.776 \pm 0.102$ & $1.011 \pm 0.132$ & $1.080 \pm 0.136$ \\
   $v_{\rm{CIC},r_{\rm{s}}=  5 \lmpc}$ & $1.067 \pm 0.108$ & $1.066 \pm 0.110$ & $1.086 \pm 0.112$ & $0.821 \pm 0.110$ & $1.069 \pm 0.142$ & $1.142 \pm 0.147$ \\ 
   $v_{\rm{CIC},r_{\rm{s}}= 10 \lmpc}$ & $1.181 \pm 0.121$ & $1.183 \pm 0.123$ & $1.206 \pm 0.126$ & $0.915 \pm 0.123$ & $1.197 \pm 0.160$ & $1.279 \pm 0.165$ \\
  \bottomrule
\end{tabular}

\medskip
Note: The results are estimated with one particular realization of CMB+Noise
sky and 50 realizations of mass scatter and miscentring effects. The
measurements without these two effects are listed in 2nd column, and results
with velocity uncertainties are listed in 3rd column.
\end{minipage}
\end{table*}

In this section, we test the ideas presented before, and assess their
performance when applied to mock kSZ observations. We will start with the
simplest case, in which a perfect cluster catalogue and velocity fields are
assumed to be known.  Then, we progressively increase the realism of the cases
we consider and include further sources of uncertainty. In all cases, we
explore different estimates of the velocity field and the two cluster
catalogues described in \S\ref{cluster}, unless otherwise stated. The main
results are summarized in Tables~\ref{tab_result1} and \ref{tab_result2} and
Fig.~\ref{beta_bias_plot}. 

We note that if our measurement of $\alpha$ is equal to the unity, our method
would provide an unbiased estimate of the relation between density and
velocity, which is captured by the $\beta$ parameter. The uncertainty in
$\alpha$ can be regarded as the accuracy with which $\beta$ is measured. 

\subsection{CMB primary anisotropies, residuals and instrument noise}

We start by considering the CMB primary anisotropies as the only source of
uncertainty in $\alpha$. We applied the procedure outlined in the previous
section to $50$ realizations of the CMB sky. The results are provided in the
4th column in Table~\ref{tab_result1} (labelled as `CMB'). The mean measured
value of $\alpha$ is $0.963 \pm 0.0046$ for the Mock-MaxBCG sample. For the 
more abundant MXXL sample is $0.977 \pm 0.0035$, which shows a similar bias
in $\alpha$ but the statistical error decreases. This is the first validation
of the performance of our matched filter approach. 

Even though the component separation procedure could provide a foreground
cleaned CMB map, some residuals are still left over as contaminations. Since our
measurements are performed at individual clusters, tSZ residuals are more likely
to add-up coherent contaminations to our results, showing up as the term
$\epsilon$ in equation~(\ref{eq:beta_est_i}). We initially presume that
$\epsilon/v_{\rm{rec}}$ goes to zero when averaging over whole cluster samples.
In order to check further how true this assumption is, we generate a tSZ map
with the same data set, and then a few per cent amplitude of the tSZ signal is
added to the CMB + kSZ map as residual. We test 1, 5 and 10 per cent tSZ amplitudes
and find that residual in term $\epsilon/v_{\rm{rec}}$ almost gone and bias up
the final $\alpha$ with less than 1, 4 and 7 per cent, respectively. So we think at
a few per cent level, residual would have little impact on $\alpha$ estimation. 

Another important source of uncertainty on small scales is the instrument noise.
As shown in fig.~E.3 of \citet{2013arXiv1303.5072P} and in
Fig.~\ref{map_spectra}, at the scale of our Mock-MaxBCG clusters (around 6
arcmin or $\ell \sim 1600$), the typical amplitude of this noise is similar to
that caused by primary CMB anisotropies. In our formulation of the matched
filter this contribution is implicitly taken into account, since we use the
power spectrum of our sky map itself as the noise term in equation~(\ref{eq:filter}). 

In order to assess the impact on this extra noise contribution on $\alpha$, we
have generated 50 independent maps of the CMB primary fluctuations plus \plk\
product SMICA-like instrument noise. We apply our matched filter approach and
show the result in the 5th column in Table~\ref{tab_result1} (labelled as
`CMB+Noise'). As expected, the mean value of $\alpha$ remains the same, since
this new noise component is uncorrelated with the signal. The associated
uncertainties, however, roughly double.

We note that the uncertainties on $\alpha$ estimated from the variance across
$50$ sky realizations and by the matched filter procedure agree remarkably well.
For instance, when only CMB sky included, the scatter on $\alpha$ from the 50
sky realizations is $0.0459$, compared with matched filter output value of
$0.0455$. For the case of `CMB+Noise', the value is $0.089$ compared with
$0.096$. A further support for our implementation of the matched filter approach can
be obtained by comparing the results provided above with those obtained from a
simple AP filter, which are provided in the 2nd and 3rd columns of
Table~\ref{tab_result1}. Even though the estimated $\alpha$ for all cases is
consistent with those obtained using a matched filter, the statistics errors
quoted are a factors of $20-30$ larger and are comparable with the level of the
signal itself. These two facts support the statistical validity and advantage
of our formulation.

In all the cases, we have considered so far there is a small bias in $\alpha$,
$\alpha \neq 1$ roughly at the $1\sigma$ level. We have checked that this
originates from the fact that the peculiar velocity measured from the kSZ is
actually a mass-weighted average over a cylinder on the sky of size
$3\times\,r_{200}$. This is not necessarily identical to the centre of mass
velocity of the cluster. This explains why the bias is slightly larger in the
MXXL cluster which contains less massive clusters. Nevertheless, we will see
that the systematic biased introduced are smaller than the statistical
uncertainty introduced by other sources of noise, and thus validates our
modelling given the accuracy with which current measurements are possible.

\subsection{The uncertainties in the cluster catalogue}

We now consider the impact of uncertainties in estimating cluster masses
observationally. Also we address the difficulty of optical cluster finders
algorithms to identify the clusters centre of mass.

\subsubsection{The mass--richness relation}
\label{mscatter}

The mass of a cluster is not a direct observable, one has to infer it from
other observed properties (e.g. optical richness, strong and weak gravitational
lensing signal, X-ray luminosity or tSZ flux signal). Although the mean
relations can be calibrated observationally or using numerical simulations,
deviations of individual clusters from the mean relation lead to a scatter on
the estimated cluster mass. Furthermore, there are other sources of scatter in
the observable--mass relationship related to line-of-sight contamination, the
dynamical state and triaxiality of the parent halo, etc. This affects the shape
of matched filter and its normalization, and therefore this introduces a
further source of uncertainty in the estimated velocity of a cluster from its
kSZ signal.  

Here, we explore this effect in the case of a optically-detected cluster
catalogue, such as the MaxBCG sample. The mass--richness relation and its
scatter for such catalogue have been studied with both observations
\citep{2007arXiv0709.1159J} and simulations
\citep{2010MNRAS.404..486H,2012MNRAS.426.2046A}. For example, 
\citet{2012MNRAS.426.2046A} give a mean relation which is described by a power
law $\langle M_{200}\rangle = M^{1.07}$, with a lognormal scatter
$\sigma_{\log_{10}(M_{200})} = 0.36$. In order to incorporate this effect in
our simulations, we assign a richness to each cluster according the following
procedure.  

We utilize the results provided by \citet{2010MNRAS.404..486H}, that the mean
mass $\langle M_{200}\rangle$, the lognormal scatter of mass
$\sigma_{\log_{10}(M_{200})}$ and cluster number density $n$ are given for
various richness $N_{200}$ bins. With these pieces of information cluster mass
distribution at each richness bin $\mathrm{pdf}(M_{200})$ can be constructed,
therefore the cluster mass function is just summation over contributions from
all richness bins,
\begin{equation}
  \dd n(M_{200})/\dd M_{200} = \sum_{i=1}^{N_\mathrm{bins}} n_i \mathrm{pdf}_i(M_{200}),
\end{equation}

\noindent then this function is normalized by the total number of clusters in
our Mock-MaxBCG catalogue. After that, we divide clusters in our catalogue into 
several different logarithmic mass bins. In each of these bins, clusters are 
assigned a richness according to the probability $\mathrm{pdf}_i(M_{200})$.

Once each cluster has a richness value, we use the mean mass--richness relation
to assign an estimate for the cluster mass. Then, we construct the corresponding
filters and repeat our analysis. The results are shown in the 4th column of
Table~\ref{tab_result2}. We find that the average $\alpha$ and its uncertainty
both vary by less than 5 per cent. This is in agreement with previous works which 
showed that the scatter in mass does not have a significant impact on
filter-recovered tSZ signal \citep{2012ApJ...757....1B,2014A&A...561A..97P}.
This is a consequence of the shape of the matched filter being weakly dependent
on cluster mass as a result the concentration and cluster size depend weakly on
halo mass. Moreover, the uncertainty on mass estimation is subdominant compared
to the other sources of uncertainties related to the CMB maps.

The above discussion is based on the hypothesis that the mean mass--richness
relation could provide unbiased mass estimations for clusters in fixed richness
bin. While the mass--richness relation is usually obtained by lensing mass
calibrations. But lensing mass estimates are very difficult, mass--richness
relations obtained by different lensing mass calibrations already show an
important discrepancy \citep{2007arXiv0709.1159J,2009ApJ...699..768R}. This
would introduce a potential systematic effect on estimating cluster mass, and
consequently would take effect when translating kSZ flux into velocity
estimation (equation~(\ref{eq:beta_est_i})). In order to examine the systematic bias of
mass--richness relation on $\alpha$, we shift the mean relation  by 10, 20,
40 per cent high (low), then we find that $\alpha$ estimation will be biased 6, 10,
18 percent low (6, 14, 34 per cent high), respectively. This may act as an important
systematic bias in our method. It is still not clear how well can different
mass--richness relations represent cluster mass and its observable. It is hard to
make a specific quotation here, so we put three bias levels for reference.

\subsubsection{The cluster miscentring}
\label{miscentering}

Another effect that may seriously hamper our efforts to get an accurate value of
$\alpha$ is the offset between the centre of mass of a cluster and the centre
estimated using its optical properties. A BCG misidentification and
astrophysical processes may both cause this so-called miscentring
\citep{2007arXiv0709.1159J,2010MNRAS.404..486H}. 

We estimate the impact of this effect by randomly selecting a fraction of
clusters, $p_{\mathrm{c}}$, to be miscentred: from 40 per cent for the lowest
richness bin down to 20 per cent for the highest richness bin, and then perturb their
centre according to

\begin{equation}
\label{eq:offset_distribution}
 \mathrm{pdf}(R_{\rm{off}})=\frac{R_{\rm{off}}}{\sigma_{\rm{off}}^2}\exp\left(-\frac{R_{\rm{off}}^2}{2 \sigma_{\rm{off}}^2}\right)\ ,
\end{equation}

\noindent where $\sigma_{\rm{off}}=0.42\lmpc$ \citep{2010MNRAS.404..486H}. This
expression describes the distribution of projected distances between the
identified centre and the centre of mass of a cluster. 

We repeat our analysis for the new centres. We find a strong decrease in the
estimated value of $\alpha$ of about 20 per cent. The reason for this is that the
incorrect cluster convolved with the matched filter results in a heavy under
estimation of the clusters mass and kSZ signal. One way to reduce the problem
is to modify the matched filter with an effective kSZ signal profile that
correctly describes the presence of a set of miscentred clusters. 

The kSZ signal of clusters in a given mass can be described as a weighted
average of the profile of correctly and incorrectly centred objects. Correctly
centred clusters have a mean kSZ profile given by equation~(\ref{eq:profile}).  The
signal of miscentred clusters is a convolution by azimuthal angle of the
offset distribution with the correctly centred profile:

\begin{equation}
  \tau_i(R | R_{\rm{off}}) = \frac{1}{2 \pi} \int_0^{2 \pi} \dd \theta\ \tau_i \left( \sqrt{R^2 + R_{\rm{off}}^2 + 2 R~R_{\rm{off}}\cos(\theta)} \right)\ .
\label{eq:conv}
\end{equation}

Assuming that the offset distribution is given by equation~(\ref{eq:offset_distribution}), 
the mean kSZ profile of miscentred cluster can be written as a average over the
distribution:

\begin{eqnarray}
\tau_i^{\mathrm{mis}}(R) = \int \dd R_{\rm{off}}\ \mathrm{pdf}(R_{\rm{off}})~\tau_i(R | R_{\rm{off}}).
\end{eqnarray}

Finally the mean kSZ profile for clusters in a given mass bin $i$ is

\begin{eqnarray}
\tau_i^{\mathrm{tot}}(R) = (1 - p_{\mathrm{c}})\,\tau_i + p_{\mathrm{c}}\,\tau_i^{\mathrm{mis}}.
\end{eqnarray}

The new profile has a core, which compensates the total integrated kSZ flux
signal with a statistically correct answer. We have repeated our analysis with
the new matched filters, and show the results in the 6th column of
Table~\ref{tab_result2}. Indeed, after this correction, we recover a
statistically unbiased estimation of $\alpha$. This is a dramatic improvement
compared to the results without considering the miscentring problem. The price
for this in an increment of about 25 per cent in the uncertainty with which we
measure the kSZ effect. We note, however, that this effect also needs to be
considered in any other interpretation of the correlation between galaxies and
the kSZ signal, and in any other quantity estimated from template fitting (e.g.
the SZ decrement).

\subsection{Peculiar velocities}

\begin{figure*}
\plotside{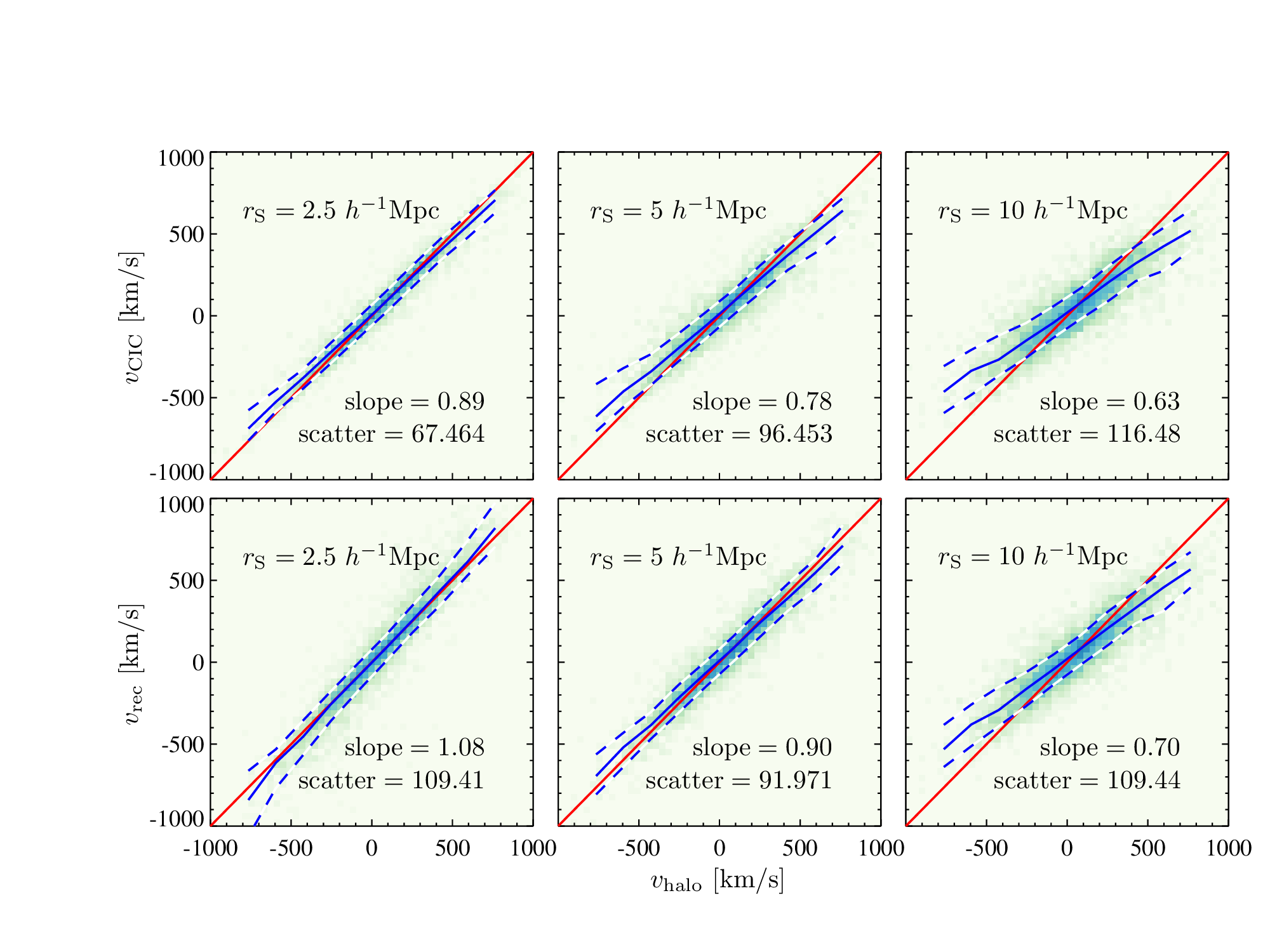}
\caption{
         Top panels: one-to-one comparison between the true radial velocities 
         of clusters $v_{\rm{halo}}$ and cluster velocities $v_{\rm{CIC}}$ from 
         CIC assignment. Bottom panels: one-to-one comparison between the true 
         radial velocities of clusters $v_{\rm{halo}}$ and cluster velocities 
         $v_{\rm{rec}}$ from reconstruction method. The red solid line in each 
         panel is the 1:1 line. The blue solid lines show the median relation
         binned by $v_{\mathrm{halo}}$. The dashed blue lines indicate the region
         containing the central 68 per cent scatter of the median relation.
         Left-hand panels on smooth scales of $2.5\lmpc$, 
         middle panels $5\lmpc$ and right-hand panels $10\lmpc$.
         }
\label{vel_comp}
\end{figure*}

When dealing with observations, the velocity of clusters is unknown, and one
needs to resort to indirect estimations. In order to assess the impact of this, 
we repeat our measurements, but now employing different estimations for the
velocity of clusters, as listed in \S3.4. The results are provided in the
$2-4$th rows and in the $5-7$th rows for velocities estimated using linear
theory and CIC interpolation, respectively.

In the case of CIC velocities, $v_{\rm{CIC}}$, we see that the $2.5\lmpc$
smoothing provides an unbiased estimate of $\alpha$, whereas the $5$ and
$10\lmpc$ smoothing overestimate its value (i.e. underestimate the clusters
velocity) by roughly $1\sigma$ and $2\sigma$. 

In the case where reconstructed velocities, $v_{\rm{rec}}$ are considered, we
see that the smallest smoothing scale returns a value for $\alpha$ between
20 and 25 per cent smaller than the unity. For larger smoothing scales, the
underestimation decreases and for the $10\lmpc$ smoothing, we recover a
(biased) value consistent with that in the case of CIC. However, uncertainties
are about 30 per cent larger in the latter case.

Now we explore these results further. In Fig.~\ref{vel_comp}, we show one-to-one
comparisons between clusters true line-of-sight velocity, and (i) the CIC
smoothed velocities (top panels) and (ii) reconstructed velocities based on
linear theory (bottom panels). As stated before, we consider three smoothing
scales, $r_{\rm{s}}=$2.5, 5 and 10$\lmpc$.

In all cases, we see a strong correlation between the true and estimated
velocities.  The scatter increases as we consider larger smoothing scales, and
also the scatter for reconstructed velocities is larger than for CIC
velocities.  Also, for both estimation methods, we see that the velocities
are systematically underestimated for large smoothing scales. This can be seen
more clearly by comparing the 1:1 relation (red line) with the blue diagonal
line, which shows the median value of velocity estimated in bins of
$v_{\mathrm{halo}}$. Note that the slope of the relation for the case of
reconstructed velocities is always steeper than the CIC counterparts, which is
a consequence of linear perturbation theory breaking down and overestimating
the divergence of the velocity field and thus of its line-of-sight component
\citep[see also fig. 7 of][]{2012MNRAS.425.2422K}.

Overall, we appreciate that the estimated velocity field is a balance between
two competing effects: (i) the accuracy of linear perturbation theory and (ii)
how well a smoothed field approximates the actual velocity of the cluster. On
small scales, we approximate better the velocity of cluster, however, linear
theory breaks down overestimating the velocity field. On large scales, the
performance of linear theory improves; however, the recovered smoothed velocity
field underestimates the velocity at the clusters position. In other words,
while the velocity field shows a high coherence, the velocity structure of
regions as small as $2.5\lmpc$ can affect systematically high-precision
measurements of the kSZ effect.

\subsubsection{The systematics from velocity reconstruction} \label{vel_uncer}

An additional source of uncertainty is introduced by the estimation of the DM 
density field from a distribution of galaxies. In particular, effects
such as survey mask, selection function, shot noise, redshift space distortions
will all add extra uncertainties in the recovered velocity field.  One example
of recovering a continuous and smooth 3D density field from a group of galaxies
is presented in \citet{2010MNRAS.409..355J}, who applied a Hamiltonian density
algorithm {\small \bf{HADES}} to SDSS data (Release 7) and returned a set of $40000$
possible realizations of the density field given the data and observational
setup. An additional complication comes from the fact that we observe the
galaxy field in redshift space, thus one needs to assume a value of $\beta$ to
estimate the corresponding velocity field. This, however, can be coupled with
the kSZ measurements to sample different values of $\beta$ in a self-consistent
manner as we measure $\alpha$.

The total error associated with the reconstructed velocity can be modelled as the
sum of two independent terms

\begin{equation}
\sigma^2_{\rm{rec}} = \sigma^2_{\rm{obs}} + \sigma^2_{\rm{meth}}. 
\end{equation}

\noindent $\sigma_{\rm{obs}}$ refers to an uncertainty that depends on the
particular observational setup propagated through the density reconstruction
method. This term varies as a function of position and distance to the
observer, and slightly depends on velocity. The second term,
$\sigma_{\rm{meth}}$, is velocity-independent term and it accounts for the
uncertainties in the method itself (i.e. the scatter shown in
Fig.~\ref{vel_comp}), which is around $100~\kms$. 

In practice, it is difficult to estimate each of these two terms independently
due to their correlation. However, the total error budget, $\sigma_{\rm{rec}}$,
can be determined. As studied in Appendix~\ref{scatter_data}, a typical value of
$\sigma_{\rm{rec}}$ is about $350~\kms$, for an SDSS-like survey. With more
accurate reconstruction methods the uncertainty in $\sigma_{\rm{meth}}$ can be
reduced to $20~\kms$ \citep{2012MNRAS.425.2422K}. Therefore,
$\sigma_{\rm{rec}}$ would be mainly determined by $\sigma_{\rm{obs}}$, which is
position and distance dependent.

The value of $\sigma_{\rm{rec}}$ inferred above clearly indicates that the
current limitation is in the quality of the data of a target galaxy survey, and
the method to infer the DM density field. This turns into another issue that how
well velocity reconstruction method could return the correct sign for each
cluster. This would potentially affect the efficiency of our estimation on
$\alpha$. But with the aim to offer a clean forecast for our method, reconstruct
velocities based on numerical simulation are used in this paper, without fully
considering observational effects. We plan to further test the sign issue with
more realistic data through the {\small \bf{HADES}} pipeline in a forthcoming
work.

In this paper, we just approximately account for the effect of
$\sigma_{\rm{rec}}$, our Mock-MaxBCG clusters are assigned uncertainties in
their line-of-sight velocity by interpolating the velocities and uncertainties
reconstructed by {\small \bf{HADES}} (shown in Appendix~\ref{scatter_data}) to
the positions of our clusters. The total uncertainty can be incorporated in our
approach by using the full form of the weights shown in equation~(\ref{eq:weight1}). 

We have repeated our analysis with this extra source of uncertainty. However,
the estimated value of $\alpha$ remained almost identical to our previous case.
This is because $\sigma_{\rm{rec}}$ plays a minor role in $w_i$ compared to
$\sigma_{\rm{kSZ}}$, which is dominated by the primary CMB fluctuations and
instrumental noise and is at the level of a few thousands $\kms$. Hence, the
quality of the reconstructed velocity field does not affect significantly the
estimated $\alpha$, but mainly the $\alpha_i$ value for each clusters.

\begin{figure*}
\plotside{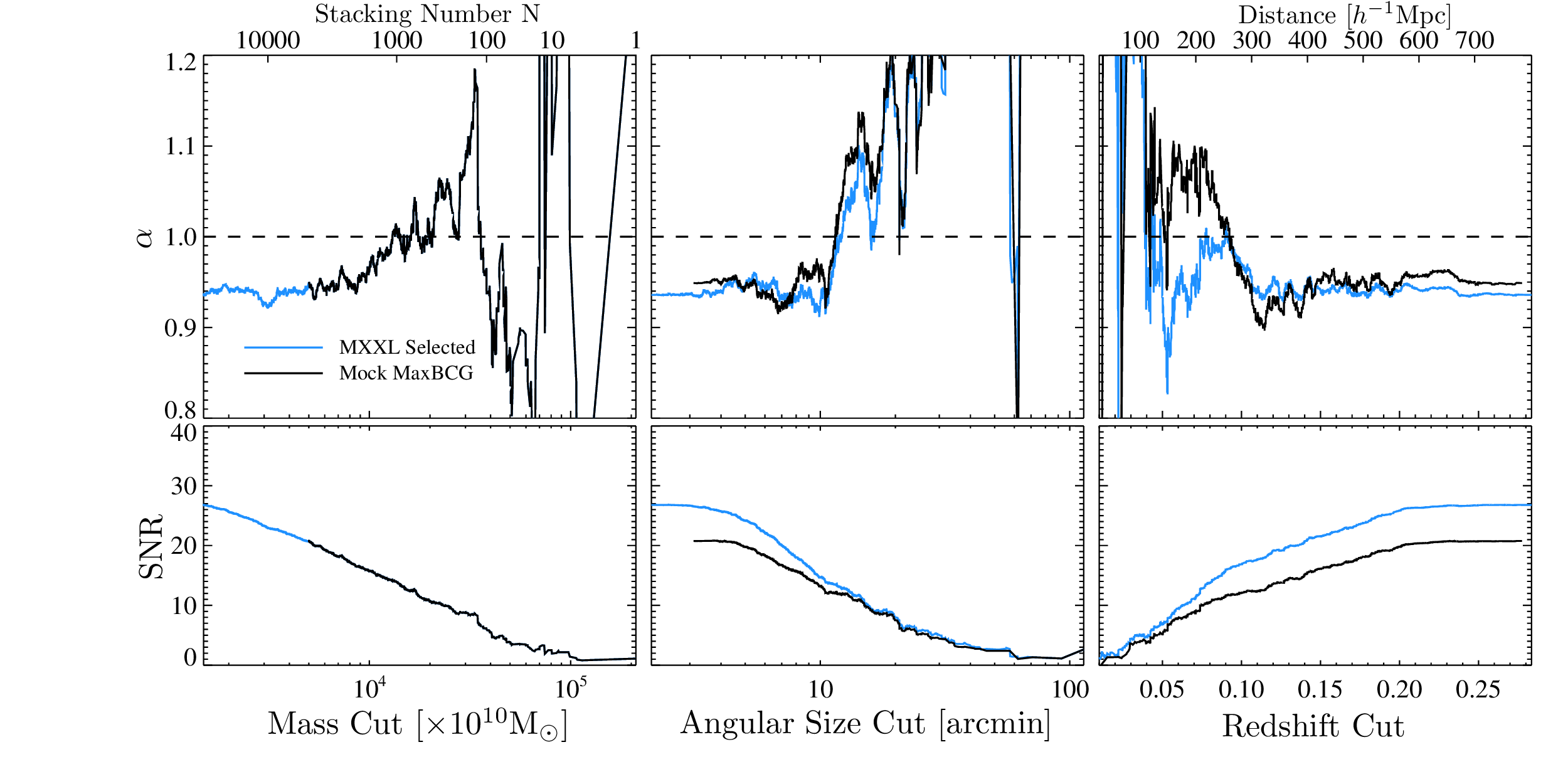} \\
\plotside{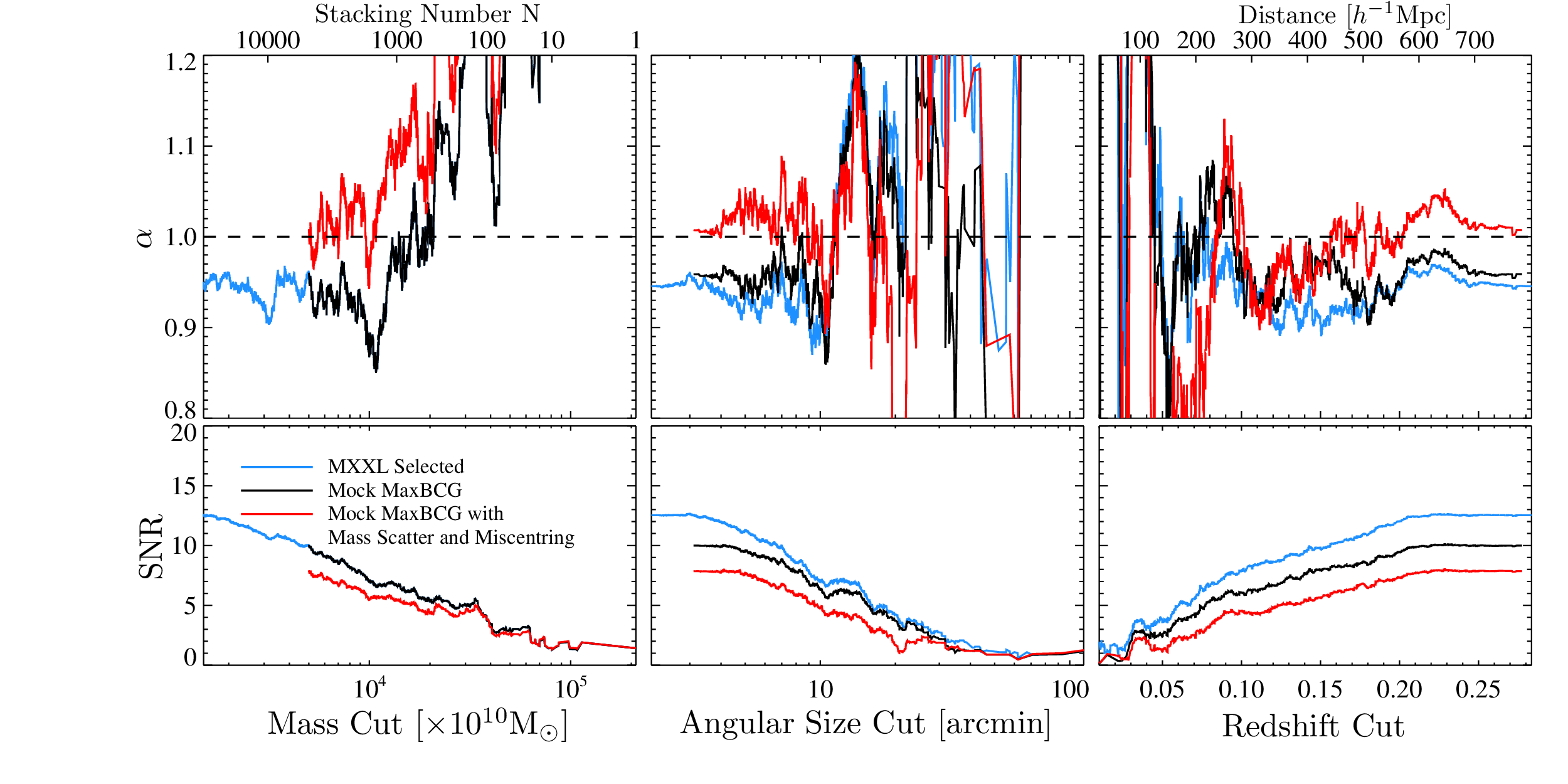}
\caption{Top row set: top panels show the weighted estimation $\alpha$
         (equation~(\ref{eq:beta_est})) as a function of mass cut (left column),
         angular size cut (middle column) and redshift cut (right column) which
         are used to test cluster selection criteria from our catalogues. The
         corresponding stacked number is shown on top axis of mass cut case.
         Bottom panels show the corresponding SNR of
         $\alpha$ following the same way. The two cluster catalogues
         are used and shown as blue (MXXL selected) and black (Mock MaxBCG)
         curves. The analysis is based on map of kSZ+CMB and $\alpha$
         using $v_{\rm{halo}}$ are shown as example. 
         Bottom row set: the same as top row set plots, but $\alpha$ is analysed
         based on map of kSZ+CMB+Noise. Results of Mock MaxBCG when considering
         mass scatter and miscentring problems are also presented with red
         curves. }
\label{stacked}
\end{figure*}


\subsection{Cluster catalogue selection}
\label{cluster_sel}

\begin{figure*}
\plotside{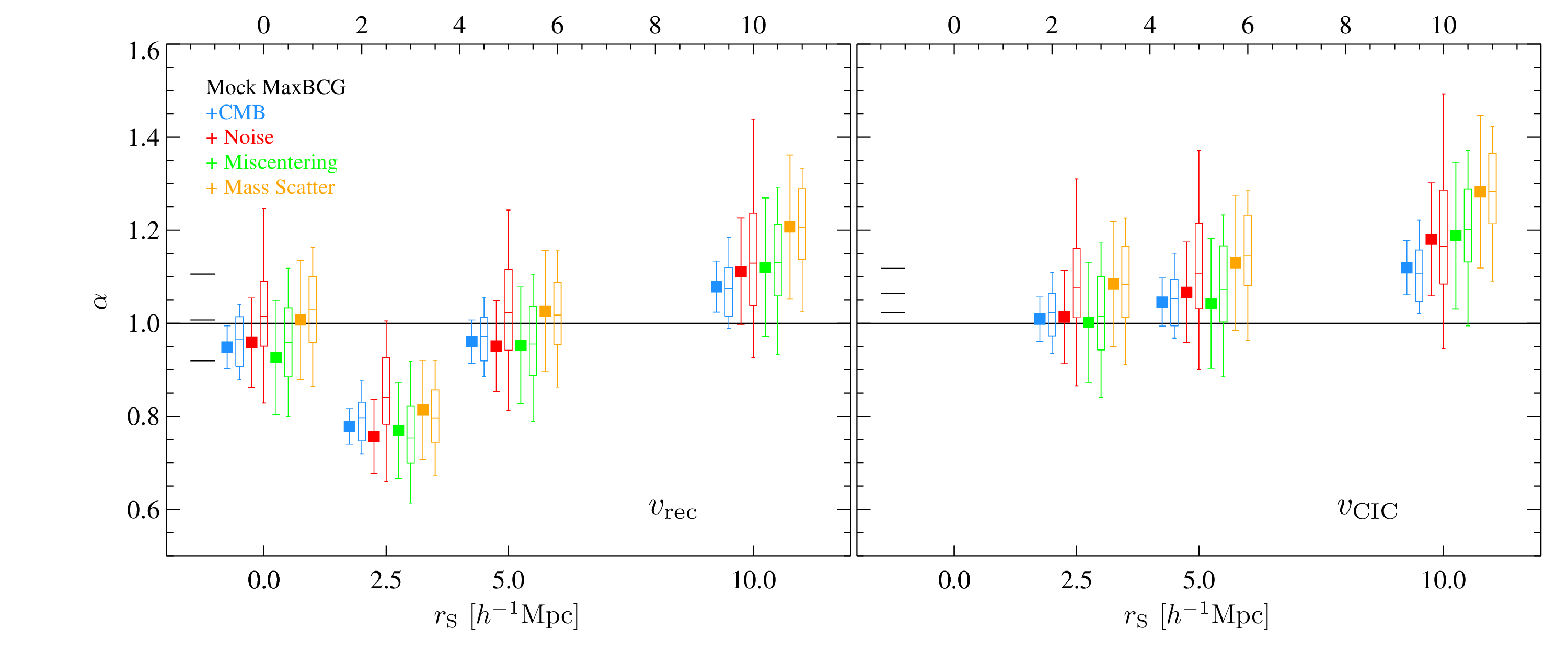}
\caption{The measurement of $\alpha$ for Mock-MaxBCG clusters with different velocity
         references. Data sets are listed according to velocity used,
         $v_{\rm{rec}}$ (left-hand panel) and $v_{\rm{CIC}}$ (right-hand panel), their
         positions according to the smooth length $r_{\mathrm{s}}$. Specially
         measurement using $v_{\rm{halo}}$ is positioned in left-hand panel with
         $x$-axes $r_{\mathrm{s}}=0$.  For each set of data points, measurements
         with continuous including noise are distinguished with colours. The
         filled squares are single measurements, next to them are the
         corresponding measurement distributions estimated over $50$
         realizations. The bars, boxes and whiskers show the median and the 1,
         16, 84 and 99 percentiles of the $\alpha$ distributions. The black
         solid lines at left-hand side of each panel present the bias level computed
         by taking median of $v_{\rm{halo}}/v_{\rm{rec}}$ or
         $v_{\rm{halo}}/v_{\rm{CIC}}$ at the three smoothing scales, from bottom
         to top 2.5, 5, 10$\lmpc$, respectively.}
\label{beta_bias_plot}
\end{figure*}

A central part of our method is the existence of an appropriate cluster
catalogue, and the accuracy of our method depends on its properties. Therefore,
in this subsection and in Fig.~\ref{stacked} we explore how the selection
criteria affect the accuracy with which we measure $\alpha$. In particular, we
consider different cuts in mass, in angular size and in redshift.  For each
threshold, we consider a simple case where we include only the CMB as a source
of noise (top panel) and another in which we consider further sources of
uncertainty (bottom panel). For clarity, we use the measured centre of mass
velocity of each cluster. The results are shown in Fig.~\ref{stacked} and we
discuss them next.

\subsubsection{Mass/Richness}

We recall that so far we have shown results for two different catalogues. (i)
`MXXL Selected', which contains all clusters in a SDSS-like volume with mass
above  $1.5\times10^{13}\msun$, and (ii) `Mock MaxBCG', which is a sub-sample of
previous catalogue with a higher mass threshold $5\times10^{13}\msun$. The
latter corresponds to a threshold of optical richness $N_{200}=10$ in the real
MaxBCG catalogue. Systems above that richness are identified at high
significance and suffer little contamination; however, below that limit there
is still information about overdensities in the Universe. Thus, we explore
whether these can increase the quality of our kSZ measurements.

In the leftmost panels of Fig.~\ref{stacked}, we show the SNR as a function of
the minimum cluster mass considered in the catalogue. The number of objects
that this implies is displayed in the top axis. First, we note that the
estimated value for $\alpha$ converges after roughly few thousands objects are
included, noting that no difference between the Mock MaxBCG and the MXXL selected
catalogues. Secondly, the SNR is roughly proportional to the number of clusters
used, but the change is not as dramatic as one would have expected.  There is a
factor of 5 difference in the number of clusters among catalogues, so if we
simply consider the number of systems, then one would have expected a reduction
of a factor of $\sqrt{5} = 2.23$. The actual SNR improvement is about 25 per cent.
This is because, the newly added systems will be much less massive, so their
associated signal and angular extent on the sky are also smaller. 

\subsubsection{Angular size}

We now consider the effect of varying the minimum angular size of clusters
included in the catalogue. We show the results in the middle panels of
Fig.~\ref{stacked} and, as in the previous case, top and bottom pales show two
cases where we consider different sources of uncertainty. Black and blue lines
indicate the results for the `Mock BCG' and `MXXL Selected' catalogues,
respectively.

We see that the value for $\alpha$ quickly converges after we include objects
with an apparent size of the sky larger than 10 arcmin. The SNR also increases
rapidly as we include smaller and smaller objects; however, there is a clear
saturation at $4-5$ arcmin, coinciding with the beam size in our simulated
\plk-like CMB skies (5 arcmin). The plateau in the SNR seems to appear more
smoothly in the bottom panel, which is because clusters comparable to, but
larger than, the beam size are already being affected by the instrumental noise
(which becomes dominant at  around $\ell \sim 1600$). Nevertheless, it is
important to note that the value of $\alpha$ is not affected and is largely
insensitive to the threshold angular size we employ. This supports again the
robustness of our approach.

Finally, the MXXL catalogue returns a higher SNR than the Mock MaxBCG, at all
angular thresholds. Combining this information with that in the previous
subsection, we see that the gain in SNR from reducing the threshold mass,
largely originates from the small but nearby systems which are well resolved
above the beam size.

\subsubsection{Redshift}

To end this section, in the rightmost panel of Fig.~\ref{stacked}, we show the
results we obtain as we vary the maximum redshift of clusters included in our
analysis. We see that the bulk of the signal originates from clusters below
redshift 0.2. This is partially because of the selection function applied to
our catalogues, but mainly because high-redshift clusters have small angular
extents, despite the enhanced volume covered. Finally, as expected, there is a
roughly constant offset between the two cluster catalogues we consider, due to
the higher number of objects in the MXXL selected catalogue, at all redshifts.

From this section, we can conclude that current cluster catalogues would capture
almost all the signal available in a CMB experiment like \plk.  Reducing the
mass threshold does not increase significantly the SNR of the measurement
because the extra systems will be less massive and thus contribute less to the
total signal, and also because a considerable fraction of them will be below
the beam size of the experiment. Moreover, the small systems usually have
inaccurate measurement of cluster properties (richness, positions and so on),
which may affect the definition of the matched filter and related signal. In the
light of this, it seems that much more gain can be found in extending the sky
coverage of survey within which the cluster population is well characterized
rather than in employing lower mass systems.

\subsection{Summary of results} \label{beta_bias}

In this section, we have explored different sources of uncertainties in our
proposed procedure. These findings are summarized in Fig.~\ref{beta_bias_plot},
where we plot our results grouped by the velocity estimation used and the
respective smoothing scales. For each set, we show measurements progressively
including four noise terms, as indicated by the legend. For each case filled
squares show the results of a single measurement with error bars given by the
matched filter formalism, whereas box plots show the median and the 1, 16, 84
and 99 percentiles of measurements over an ensemble of $50$ realizations.

In the following, we summarize the most relevant findings: \begin{itemize}

\item[1)] The main sources of statistical errors are: CMB primary fluctuations,
instrumental noise, mass estimation, and cluster miscentring, each of which
contributes about 13, 42, 5, and 40 per cent of the total error variance. 

\item[2)] The main source of systematic errors is the estimation of velocity
fields. There is a compromise between a smoothing scale small enough, such
that it captures accurately the features of the velocity field, and a scale
large enough such that perturbation theory is accurate. Additionally, if the
miscentring is not properly accounted for, then there is a bias in the
measurements of about 20 per cent.  

\item[3)] Another potential source of systematic effect is from the mean
mass--richness relation. If mass--richness relation provide a mass estimation
biased by 10, 20, 40 per cent high (low), then $\alpha$ estimation will be biased
6, 10, 18 per cent low (6, 14, 34 per cent high), respectively.

\item[4)] Currently, the velocity field can be reconstructed to high accuracy
using linear or higher order perturbation theory \citep{2012MNRAS.425.2443K}.
However, a source of uncertainty that remains dominant is the transformation
from galaxy to DM overdensities.

\item[5)] A MaxBCG-like cluster catalogue includes most of the available
signal.  Smaller systems do not increase the SNR substantially, due to
their small angular sizes and weak intrinsic signals. However, broader sky
coverage would lead to a considerable gain in the SNR.

\item[6)] For a \plk-like experiment and a MaxBCG-like cluster
  catalogue, we forecast a $7.7\sigma$ measurement of the kSZ,
  assuming an estimate for the velocity field using linear theory and
  a $5\lmpc$ smoothing scale. Alternatively, this measurement can be
  interpreted as 13 per cent constraint on the value of $\beta$.
\end{itemize} 

Despite the uncertainties related to the cluster catalogues, we have shown the
potential that exists in the kSZ effect to measure cosmic velocity fields and thus
to place constraints that complement those from other cosmological probes.

\section{Conclusions}
\label{disc_conc}

In this paper, we have proposed and investigated a scheme to measure
the kSZ effect with relatively high signal-to-noise. The method
combines the matched filter approach, an independent catalogue of
clusters, and the velocity field predicted by perturbation theory
applied to a galaxy redshift survey. The results can be used to
explore the properties of ionized gas in clusters or to constrain the
value of $\beta=f(\Omega_\mathrm{m})/b$. The latter, in turn, can be used to
place constraints on the gravity law that connects the cosmic density
and velocity fields.

We have shown the efficiency and accuracy of our approach by applying
it to mock CMB maps, which contain a realistic kSZ signal as predicted
by a large cosmological $N$-body simulation. Using a cluster catalogue
similar to those extracted from the SDSS data, Planck-like CMB maps,
and an estimate for the velocity field based on linear theory, we
forecast a $7.7\sigma$ detection of the kSZ effect. This result
includes the effect of several sources of uncertainty: primary CMB
fluctuations, instrumental noise, mass scatter, and cluster
miscentring.  Each of these effects is responsible for 13, 42,
5, 40 per cent of the total error variance, respectively. In addition, we
highlighted that if the potential miscentring of clusters is not
taken into account properly, a bias of about 20 per cent is induced in the
recovered signal. Similarly, if the scale on which the velocity field
is reconstructed is too small, then perturbation theory is inaccurate,
whereas if it is too large, then the features of the velocity field
are not properly resolved. Unless corrected, both effects introduce
systematic errors in kSZ estimates.

We also explored how the accuracy of our method depends on details of
the cluster catalogue. For the cases we considered, the typical
angular size of clusters corresponds to the scale on which the effect
of instrumental noise and beam size become important for a Planck-like
CMB experiment.  This implies that the kSZ signal of clusters with
small angular sizes will be smeared out, reducing their contribution
to the total accuracy of the detection. A similar effect is present
when we varied the range of redshift and mass of clusters included in
our catalogue. It appears that current cluster catalogues would
capture most of the signal available, since lower mass systems do not
significantly increase the SNR of the measurement. On the other hand,
broader sky-coverage would lead to improved constraints.

Despite the realism of the mock skies adopted throughout this work,
there are several effects which we have neglected. Most notable is the
impact of hydrodynamical interactions on the kSZ signal. For instance,
feedback from supermassive black holes at the centre of massive
galaxies can alter the distribution of mass inside clusters and,
potentially, even expel gas from the cluster altogether. However,
these effects are still highly uncertain and it is unclear that they
would be large enough to significantly alter the kSZ signal.  Once
understood, such effects could easily be incorporated in our formalism
through a modified model for the signal profile. By dividing the kSZ
measurements according to cluster mass, such systematic effects could
be detected through an apparent dependence of the cosmological signal
on cluster mass.


\section*{Acknowledgement}
ML thanks Carlos Hern\'{a}ndez-Monteagudo, Stefan Hilbert, Jun Pan and Cheng Li
for helpful discussions. SW and RA acknowledge support from ERC Advanced Grant
246797 ``GALFORMOD''. JJ is partially supported by a Feodor Lynen Fellowship
by the Alexander von Humboldt foundation and Benjamin Wandelt's Chaire
d'Excellence from the Agence Nationale de la Recherche. The authors would also
like to thank the anonymous referee for helping improve this paper.



\appendix

\section{Behaviour of velocity uncertainties with true data at positions of MaxBCG clusters}
\label{scatter_data}

\begin{figure*}
\plotside{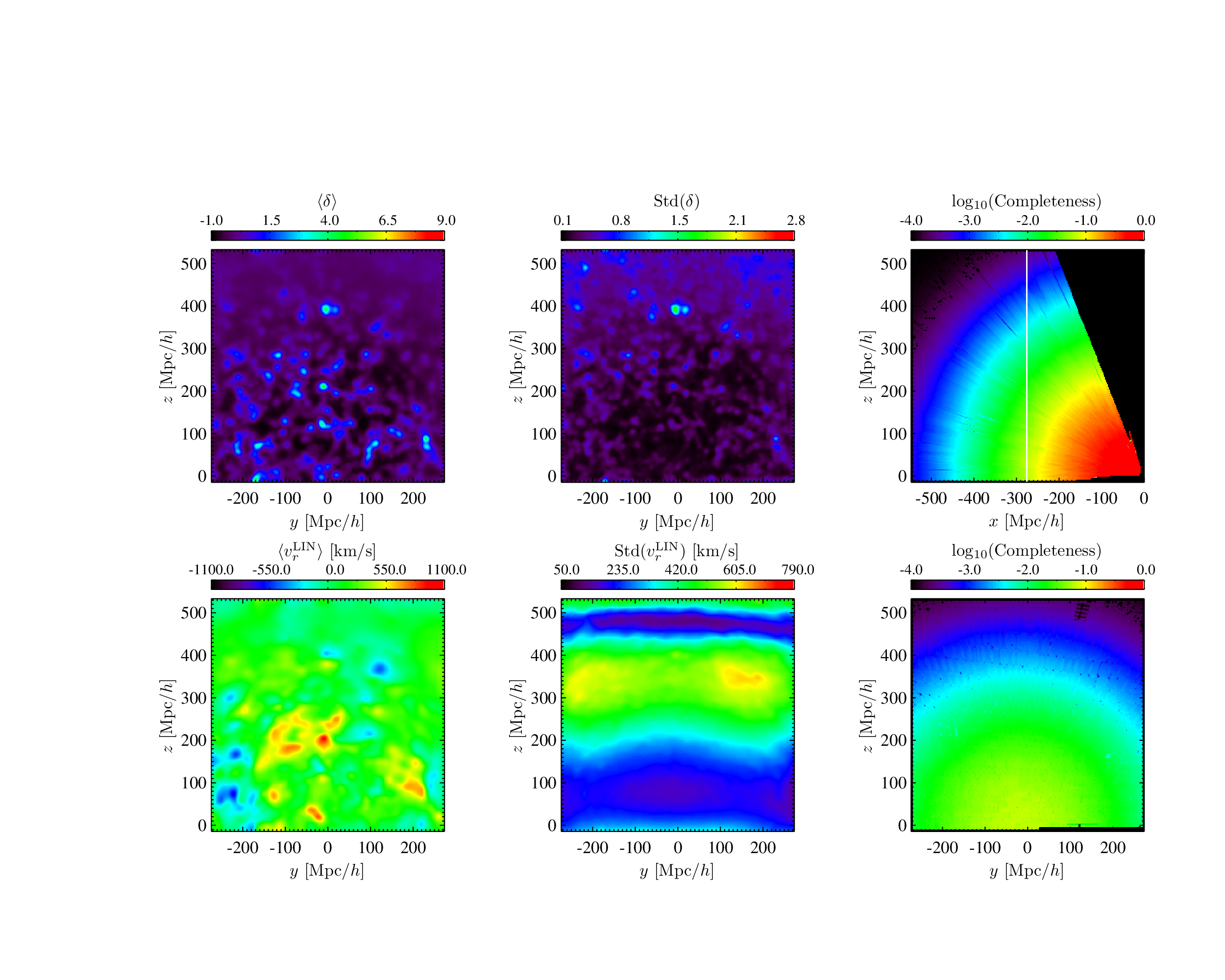}
\caption{
         Slices for ensemble (4000 realizations) mean (left column) and standard
         deviation (middle column) of density contrast $\delta$ (top panels) and
         radial velocity $v_\mathrm{r}^{\mathrm{LIN}}$ (bottom panels).  The right
         column shows a slice through the completeness function along $y$-axes and
         $x$-axes, respectively. The white solid line indicates the $x$-axes position
         of all other slices.
         }
\label{rec_vel1}
\end{figure*}

\begin{figure*}
\plotside{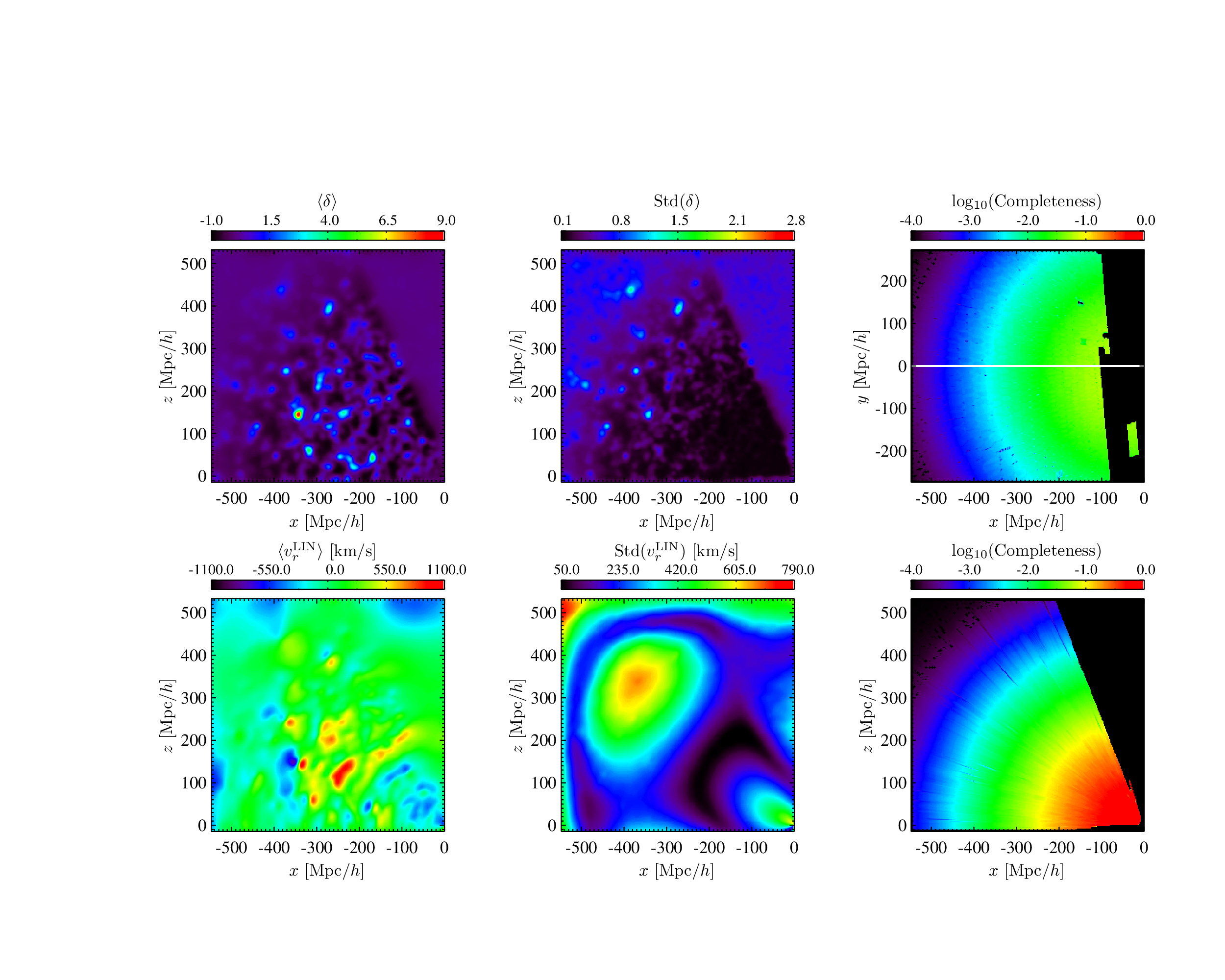}
\caption{
         Same as Fig.~\ref{rec_vel1} but density contrast $\delta$ (top panels)
         and radial velocity $v_\mathrm{r}^{\mathrm{LIN}}$ (bottom panels) are shown
         along the $y$-axes. The right column shows a slice through the
         completeness function along $z$-axes and $y$-axes, respectively. The white
         solid line indicates the $y$-axes position of all other slices.
         }
\label{rec_vel2}
\end{figure*}

In this work we use a sub-sample of 4000 density field realizations previously
generated by the {\small \bf{HADES}} (HAmiltonian Density Estimation and
Sampling) algorithm \citep{2010MNRAS.409..355J}. The {\small \bf{HADES}}
algorithm is a full scale Bayesian inference framework providing detailed
reconstructions of the 3D density field from galaxy redshift surveys and
corresponding uncertainty quantification by exploring a highly non-Gaussian and
non-linear lognormal Poissonian posterior via efficient implementations of a
Hybrid Monte Carlo method \citep[][]{2010MNRAS.407...29J}. As a result, this
algorithm provides a numerical representation of the target posterior
distribution, in terms of density field realizations constrained by
observations, permitting to thoroughly propagate uncertainties to any finally
inferred quantity. In the following, we build upon the results obtained by
\citet{2010MNRAS.409..355J}, which provide realizations of constrained density
fields in a cubic Cartesian box of side length $547.5\lmpc$ and $256^3$ voxels
inferred from the SDSS DR7 main sample \citep[][]{2009ApJS..182..543A}. The
lower-left corner of the volume locates at $[-547.5,-273.75,-14.6] \lmpc$ and 
the
observer is placed at $[0,0,0]$. To compute linear velocity fields, we smooth
these density fields on length-scales of $5\lmpc$  and apply
equation~(\ref{eq:vel_lin}).  Subsequently,  we project the ensemble of resulting 3D
velocity fields at each voxel on the observers line of sight. Given this
ensemble, mean and standard deviation of radial velocities are calculated for
each voxel. 

Results of these calculations as well as a slice through the corresponding
completeness function of the underlying SDSS survey are presented in
Figs.~\ref{rec_vel1} and \ref{rec_vel2}. As can be seen, close to the 
observer, structures are more
clearly visible, while for poorly observed regions at large distances, the
ensemble mean of the density contrast drops to cosmic mean. This reflects, the
signal-to-noise properties of the underlying survey, as uncertainties increase
with distance to the observer due to selection effects.  This effect is clearly
represented by the slices through estimated ensemble means and standard
deviations as shown in Fig.~\ref{rec_vel1}. Slices through estimated ensemble means
and standard deviations for radial velocities are presented in the bottom row of
Fig.~\ref{rec_vel1}. For the mean velocity field, large speed regions coincide
with high-density regions, being least affected by observational noise. On the
contrary, ensemble standard deviation maps are more complicated to interpret.
As can be seen, even at central regions where the observational completeness is
at median level, ensemble standard deviations ranges around $400~\kms$. The
reason for this may resort in the fact, that velocities, as estimated by
equation~(\ref{eq:vel_lin}), are most sensitive to the largest scales of the cosmic
matter distributions, which are only poorly constrained by underlying galaxy
observations, due to survey geometries. Observational uncertainties on these
large scales are nevertheless correctly treated by the statistical nature of our
approach. Additionally one may worry about periodic boundary conditions, assumed
implicitly when estimating velocities via Fourier methods, which may influence
the inference of velocities. This can be overcome by carrying out fast fourier
transforms over a much larger volume, zero-padding the unobserved region.

\begin{figure*}
\plotside{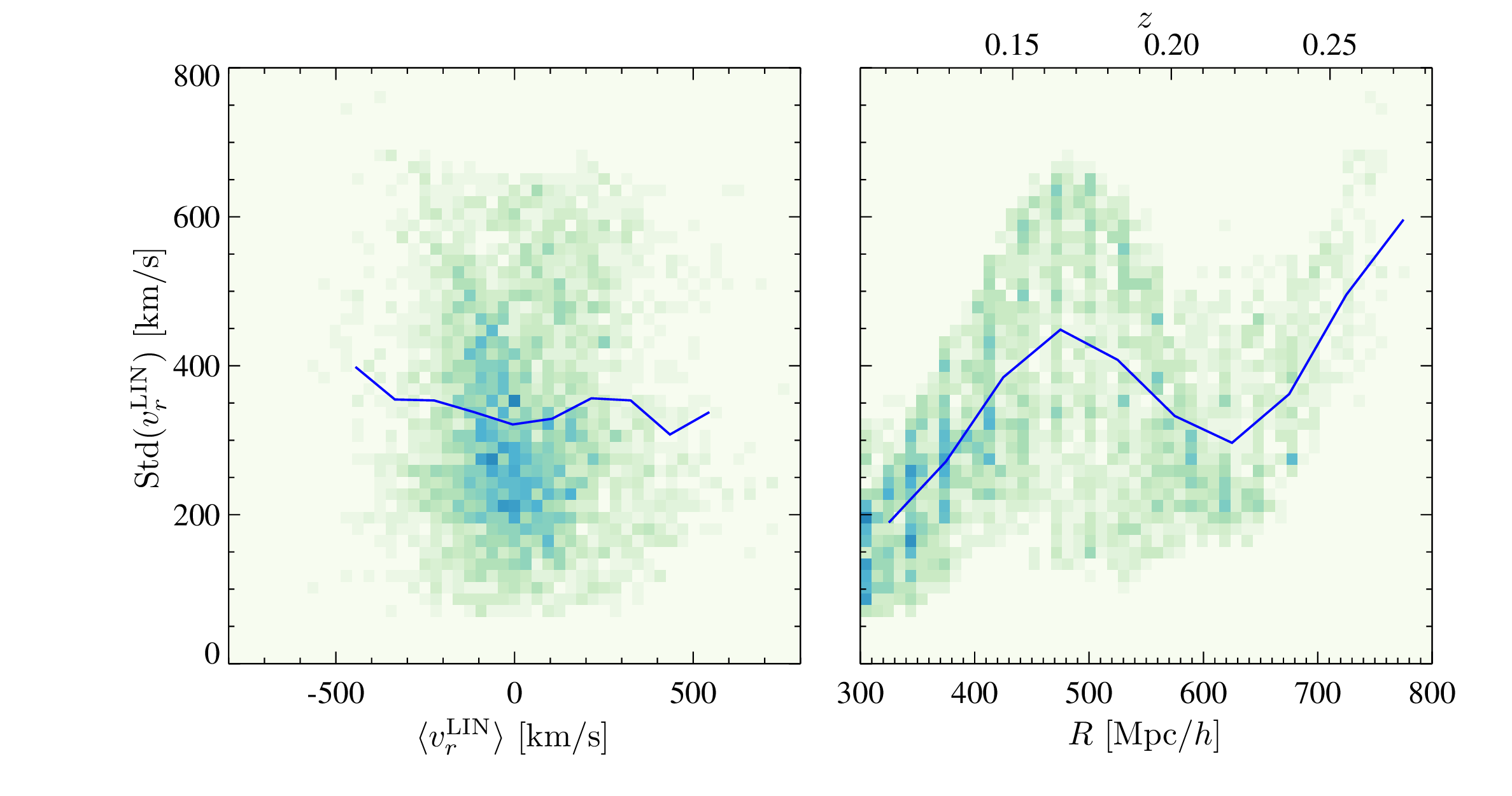}
\caption{
         Ensemble (4000 realizations) standard deviation of radial velocities
         $\mathrm{Std}(v_\mathrm{r}^{\mathrm{LIN}})$ at positions of MaxBCG clusters as
         a function of their ensemble (4000 realizations) mean radial velocities
         $\langle v_\mathrm{r}^{\mathrm{LIN}}\rangle$ (left-hand panel) and radial distances
         to the observer $R$ (right-hand panel). The intensity of background
         2D histogram is proportional to the number of clusters that reside in
         corresponding region of the plot. The blue solid line is the mean
         relation binned by $\langle v_\mathrm{r}^{\mathrm{LIN}}\rangle$ and $R$.
        }
\label{rec_vel_std}
\end{figure*}

The statistical study of the full volume is useful for the general analysis of
the reconstruction method and the goodness of density fields. As a final step,
we interpolate the reconstructed 3D velocities to positions of MaxBCG clusters
(4044 clusters reside in our reconstruction volume) and estimate ensemble means
and standard deviations of radial velocities. This addresses the issue of
velocity uncertainties inherent to such reconstructions. We check the dependence
of standard deviations on velocities and distances to the observer as
demonstrated in Fig.~\ref{rec_vel_std}. As can be seen, the standard deviation
depends weakly on ensemble mean of $v_\mathrm{r}^{\mathrm{LIN}}$. Typically, the
difference is less then $100~\kms$. The dependence on distance to the observer
is essentially strong, and standard deviation peaks at around 480 and
780$\lmpc$. The former peak shows the same complex behaviour as shown in
Fig.~\ref{rec_vel1}. The latter one is mainly due to the fact of low
completeness at such distances, and periodic boundary effects may also
contribute.  In general, velocity uncertainties are caused by the completeness
function, indicating how much information the data provides.  For the analysis
in Section~\ref{vel_uncer}, we choose a constant velocity standard deviation to be
$350~\kms$ for all clusters in our mock catalogue.  

\label{lastpage}

\end{document}